\documentclass[fleqn,10pt]{SelfArx} 

\usepackage[english]{babel} 

\usepackage{lipsum} 

  \newcommand{\changecolor}{\textcolor{black}}

\definecolor{color1}{RGB}{0,0,90} 
\definecolor{color2}{RGB}{0,20,20} 

\usepackage{hyperref} 
\hypersetup{hidelinks,colorlinks,breaklinks=true,urlcolor=color2,citecolor=color1,linkcolor=color1,bookmarksopen=false,pdftitle={Title},pdfauthor={Author}}

\JournalInfo{2017} 
\Archive{} 

\PaperTitle{Noise Control for DNA Computing} 

\Authors{Tomislav Plesa\textsuperscript{1}, Konstantinos C. Zygalakis\textsuperscript{2}, David F. Anderson\textsuperscript{3}, Radek Erban\textsuperscript{1}} 
\affiliation{\textsuperscript{1}\textit{Mathematical Institute, University of Oxford, Andrew Wiles Building, Radcliffe Observatory Quarter, Woodstock Road, Oxford, OX2 6GG, UK}} 
\affiliation{\textsuperscript{2}\textit{School of Mathematics, The University of Edinburgh, Maxwell Building, Peter Guthrie Tait Road, Edinburgh, EH9 3FD}} 
\affiliation{\textsuperscript{3}\textit{Department of Mathematics, University of Wisconsin--Madison, 480 Lincoln Drive, Madison Wi, 53706-1388}} 

\newcommand{\keywordname}{Keywords} 

\Abstract{Synthetic biology is a growing interdisciplinary field, with far-reaching applications, which aims to design biochemical systems that behave in a desired manner. With the advancement of strand-displacement DNA computing, a large class of abstract biochemical networks may be physically realized using DNA molecules. Methods for systematic design of the abstract systems with prescribed behaviors have been predominantly developed at the (less-detailed) deterministic level. However, stochastic effects, neglected at the deterministic level, are increasingly found to play an important role in biochemistry. In such circumstances, methods for controlling the intrinsic noise in the system are necessary for a successful network design at the (more-detailed) stochastic level. To bridge the gap, the \emph{noise-control algorithm} for designing biochemical networks is developed in this paper. The algorithm structurally modifies any given reaction network under mass-action kinetics, in such a way that (i) controllable state-dependent noise is introduced into the stochastic dynamics, while (ii) the deterministic dynamics are preserved. The capabilities of the algorithm are demonstrated on a production-decay reaction system, and on an exotic system displaying bistability. For the production-decay system, it is shown that the algorithm may be used to redesign the network to achieve noise-induced multistability. For the exotic system, the algorithm is used to redesign the network to control the stochastic switching, and achieve noise-induced oscillations.}

\begin{document}

\flushbottom 

\maketitle 


\thispagestyle{empty} 


\section{Introduction} 
Synthetic biology is an interdisciplinary field of science and engineering that aims to construct biochemical systems with prescribed behaviors~\cite{SynthBio1,SynthBio2}. At the theoretical level, the synthetic systems may significantly enhance our understanding of biology. At the practical level, they may have broad applications, e.g. in medicine~\cite{SynthBioMed1,SynthBioMed2,SynthBioMed3,SynthBioMed4,SynthBioMed5,SynthBioMed6}, industry~\cite{SynthBioInd1,SynthBioInd2}, and nanotechnology~\cite{NanoTech1,NanoTech2}. The systems may also be of interest to NASA for optimizing extraterrestrial explorations~\cite{SynthBioSpace}. A proof-of-concept for synthetic biology is a synthetic oscillator called the repressilator, which was implemented in vivo~\cite{SynthBio3}. The experimental advances since the repressilator range from isolated synthetic biochemical networks, to microorganisms containing partially, or even fully, synthetic DNA molecules (synthetic life)~\cite{SynthBioLife1,SynthBioLife2,SynthBioLife3,SynthBioLife4}. Examples include microorganisms  containing a synthetic bistable switch~\cite{SynthBio4}, and a cell-density controlling quorum sensor~\cite{SynthBio5}, microorganisms producing antimalarial drugs~\cite{SynthBioMed5,SynthBioMed6}, and synthetic systems designed for tumor detection, diagnosis and adaptive drug-response~\cite{SynthBioMed2,SynthBioMed3}.

The construction of biochemical networks in synthetic biology may be broken down into two steps: firstly, an abstract system is constructed, displaying prescribed properties, and taking the form of a chemical reaction network~\cite{Me,Me2,Janos1}. Secondly, the abstract network is mapped to a suitable physical network, which may then be integrated into a desired environment (e.g. a test-tube, or a living cell)~\cite{DNAComputing1}.   

In the first step of network construction, the goal is to obtain an abstract network with desired dynamics. In this paper, we consider two dynamical models of reaction networks under mass-action kinetics~\cite{David,Janos1}: the deterministic model, and the stochastic model (see Methods for more details). The deterministic model takes the form of the reaction rate equations, which are ordinary-differential equations governing the time-evolution of the species concentrations~\cite{Janos1,David}. The stochastic model takes the form of a Markov chain, which may be simulated using the Gillespie stochastic simulation algorithm~\cite{GillespieBook}. The Gillespie algorithm generates noisy copy-number time-series, with the copy-number distribution matching that obtained from the underlying chemical master equation~\cite{Janos1,David,GillespieBook,VanKampen}. The stochastic model is more-detailed, taking into an account the discreteness of the species counts, and the stochastic nature of the dynamics, which may be particularly important in biochemistry, where reaction networks may contain low-abundance species~\cite{Noise1,Noise2,SynthBio3,SynthBio4,Me2,Radek1,Radek2,Radek3}. On the other hand, the deterministic model is less-detailed, and more appropriate when the species are in high-abundance, and the discreteness and stochasticity are negligible~\cite{Convergence1}.

In the second step of network construction, the goal is to engineer a physical network whose dynamics match well the dynamics of a given abstract network, over a suitable time-interval. Engineering an appropriate physical network may proceed indirectly, by altering a preexisting physical network, or directly, by engineering a  network, which involves a given set of physical species, from scratch. The advantage of the former approach is that a preexisting network may display (partially) desirable dynamical properties. However, such a network may involve DNA and RNA molecules, proteins, and metabolites~\cite{SynthBio2}, some of which may have complex biophysical properties. Consequently, the disadvantage is that the structure (and, thus, the dynamics) of such a network cannot generally be modified in an arbitrary manner. In the latter approach, one may choose the physical species, at the expense of having to build a network from scratch. In the subfield of DNA computing, the latter approach is followed, and physical networks are engineered with chemical species consisting exclusively of DNA molecules, interacting via the toehold-mediated DNA strand-displacement mechanism~\cite{DNAComputing1}. DNA production is systematic and cost-effective, and, due to the fact that DNA biophysics is relatively well-understood, one has more freedom in controlling the structure of corresponding physical networks. More precisely, an abstract network under mass-action kinetics may be mapped to a DNA-based physical network provided it consists of up to second-order reactions, with rate coefficients varying over up to six orders of magnitude. The resulting physical network has identical deterministic dynamics as the abstract network (in the asymptotic limit of some of the kinetic parameters~\cite{DNAComputing1}), up to a scaling of the dependent variables.  A proof-of-concept for DNA computing is a synthetic oscillator called the displacillator, which was implemented in vitro~\cite{DNAComputing2}.

While the deterministic model of reaction networks is less-detailed, it is also simpler than the stochastic model, making it attractive for guiding the construction of networks, predicting accurately their mean-field behavior~\cite{DNAComputing1,SynthBio3,SynthBio4,Me,Me2,Janos1}. However, when noise is an important part of the dynamics, the stochastic model has to be considered. The intrinsic noise, often arising in biochemistry, may be controlled in two ways: it may be decreased (e.g. as in~\cite{Noise2}), in order to reduce the differences between the stochastic and deterministic dynamics. On the other hand, it may be increased, in a state-dependent manner, in order to favorably change the stochastic dynamics. In the language of molecular computing, the latter approach corresponds to exploiting the proven computational power of the stochastic reaction networks~\cite{Computing5}, by reprogramming the underlying intrinsic noise. Let us note that exploitations of the noise for enhancing biological functions have been reported in applications~\cite{Noise1,Radek3}. In this paper, we follow the latter approach, and present the noise-control algorithm (given as Algorithm~\ref{tab:thealgorithm}) which maps an input reaction network to output networks whose stochastic dynamics have an additional controllable state-dependent noise. Importantly, the input and output networks have identical deterministic model in appropriate limits of some of the parameters introduced by the algorithm. The algorithm may play a significant role in the biochemical network synthesis, allowing for a deterministic-stochastic hybrid approach. More precisely, when constructing abstract and physical networks, one may use the deterministic model to guide the construction, and then apply the algorithm to favorably modify the intrinsic noise in the stochastic model, while preserving the desired deterministic dynamics. The algorithm may also be used to adjust the intrinsic noise to favorably interact with environment-induced effects (e.g. extrinsic noise).

The rest of the paper is organized as follows: we introduce Algorithm~\ref{tab:thealgorithm} by applying it to the test network~(\ref{eq:inputnet}), which at the deterministic level displays a globally attracting equilibrium point. We show that the algorithm can favorably modify the stationary probability distribution underlying~(\ref{eq:inputnet}) at arbitrary points of the state-space, without influencing the deterministic dynamics. For example, it is shown that the algorithm may be used to redesign~(\ref{eq:inputnet}) to achieve noise-induced multimodality (multistability). We then apply Algorithm~\ref{tab:thealgorithm} to the exotic network~(\ref{eq:inputnet2}), which at the deterministic level displays a bistability involving an equilibrium point and a limit cycle. The algorithm is used to redesign~(\ref{eq:inputnet2}) to increase the stochastic switching between the two attractors, and to achieve noise-induced oscillations.


\section{A One-species Regular System} \label{sec:example1}
Consider the one-species production-decay reaction network $\hat{\mathcal{R}}(s)$, given by~(\ref{eq:inputnet}).

\noindent\begin{minipage}{.5\linewidth}
\begin{align}
\hat{\mathcal{R}}(s): \; \; \; \; \; \;   \varnothing \xrightarrow[]{k_1}  s, \nonumber \\
\; \; \; \; \; \;   s \xrightarrow[]{k_2}  \varnothing, \label{eq:inputnet}
\end{align}
\end{minipage}
\begin{minipage}{.4\linewidth}
\begin{align}
    \frac{\mathrm{d} \hat{x}}{\mathrm{d} t}  & =  k_1 - k_2 \hat{x}, \nonumber \\
		\hat{x}(0) & = \hat{x}_0. \label{eq:inputODEs}
\end{align}
\end{minipage} 

\noindent Species $s$ from network~(\ref{eq:inputnet}) reacts according to the two reactions with rate coefficients $k_1,k_2 \in \mathbb{R}_{\ge}$,  where $\mathbb{R}_{\ge}$ is the set of nonnegative real numbers, and $\varnothing$ is the zero-species (denoting species which are not of interest). In this paper, we assume reaction networks are under mass-action kinetics, with the reactions taking place in unit-volume reactors. Let us denote the concentration of species $s$ from~(\ref{eq:inputnet}) at time $t \in \mathbb{R}_{\ge}$ by $\hat{x} = \hat{x}(t) \in \mathbb{R}_{\ge}$. The initial value problem for the deterministic model (also called the drift) for network~(\ref{eq:inputnet}) is given by system~(\ref{eq:inputODEs}), with $\hat{x}_0 \ge 0$ (see also Methods). Since the deterministic model~(\ref{eq:inputODEs}) has a globally attracting equilibrium point, given by $k_1/k_2$, network~(\ref{eq:inputnet}) is said to be regular~\cite{Janos1}.

Let us denote the copy-number of species $s$ from~(\ref{eq:inputnet}) at time $t \ge 0$ by $\hat{X}(t) \in \mathbb{N}_0$, where $\mathbb{N}_0$ is the set of integers. Under the stochastic model, $\hat{X}(t)$ is modelled as a continuous-time, discrete-space Markov chain (see also Methods), which can be generated by using the Gillespie stochastic simulation algorithm~\cite{GillespieBook}. Given $\hat{X}(t)$, there will be a mean interevent time until one of the reactions from~(\ref{eq:inputnet}) fires. The mean interevent time is given by $1/\hat{\alpha}(\hat{X}(t))$, and when the event takes place, the probability that the $i$-th reaction from~(\ref{eq:inputnet}) fires is equal to $\hat{\alpha}_i(\hat{X}(t))/\hat{\alpha}(\hat{X}(t))$, for $i \in \{1,2\}$. Here, $\hat{\alpha}_1 = k_1$, and $\hat{\alpha}_2(x) = k_2 x$, are the so-called propensity functions of the first, and second, reactions from~(\ref{eq:inputnet}), respectively. Function $\hat{\alpha}(x) = k_1 + k_2 x$ is the total propensity function of network~(\ref{eq:inputnet}), i.e. the sum of propensity functions of all the underlying reactions.

We now wish to structurally modify network~(\ref{eq:inputnet}) in such a way that the deterministic model from~(\ref{eq:inputODEs}) is preserved, while an arbitrary nonnegative function, defined on a bounded discrete domain, is added to the total propensity function of~(\ref{eq:inputnet}). The latter requirement implies that the interevent time would be controllably decreased in a state-dependent manner. Equivalently, the two requirements imply that a controllable state-dependent noise would be introduced into the stochastic dynamics. We have designed a three-step algorithm, given as Algorithm~\ref{tab:thealgorithm}, which achieves such goals for arbitrary reaction networks under mass-action kinetics. Let us describe properties of the algorithm by applying it on network~(\ref{eq:inputnet}). 

Firstly, we wish to introduce an additional species $\bar{s}$ into network~(\ref{eq:inputnet}), in such a way that species $s$ and $\bar{s}$ satisfy a pairwise stoichiometric conservation law. Secondly, we require that the enlarged network has the same deterministic model as network~(\ref{eq:inputnet}), despite the added species $\bar{s}$, which may be achieved by adding another auxiliary species. More precisely, let us consider network $\hat{\mathcal{R}}^1(s,\bar{s}) \cup \mathcal{R}_1^2(\bar{s})$, given by:
\begin{align}
\hat{\mathcal{R}}^1(s,\bar{s}): \;  &  & \bar{s} + I^1 & \xrightarrow[]{k_1}  s + I^1, \nonumber \\
  &  & s & \xrightarrow[]{k_2}  \bar{s}, \nonumber \\
\mathcal{R}_1^2(\bar{s}): \;  &  &  \varnothing & \xrightarrow[]{1/\mu}  I^1, \nonumber \\
  &  &  \bar{s} + I^1 & \xrightarrow[]{1/\mu}  \bar{s}. \label{eq:outputnet}
\end{align}
Species $s, \bar{s}, I^1$ from~(\ref{eq:outputnet}) react according to the four reactions with rate coefficients $k_1, k_2, 1/\mu \in \mathbb{R}_{\ge}$. Network $\hat{\mathcal{R}}^1 = \hat{\mathcal{R}}^1(s,\bar{s})$, given in~(\ref{eq:outputnet}), is obtained from network $\hat{\mathcal{R}} = \hat{\mathcal{R}}(s)$, given by~(\ref{eq:inputnet}), in the following way: since the first reaction in $\hat{\mathcal{R}}$ \emph{increases} copy-number of $s$ by one, $\bar{s}$ and $I^1$ are added to the reactants of the reaction, and $I^1$ is added to the products, leading to the first reaction in $\hat{\mathcal{R}}^1$. Since the second reaction in $\hat{\mathcal{R}}$ \emph{decreases} copy-number of $s$ by one, $\bar{s}$ is added to the products, leading to the second reaction in $\hat{\mathcal{R}}^1$. This ensures that the desired conservation law holds. The superscript in $I^1$ indicates that species $I^1$ is involved as a catalyst in a reaction of $\hat{\mathcal{R}}^1$ in which $s$ is \emph{increased by one}. The subscript in $\mathcal{R}_1^2 = \mathcal{R}_1^2(\bar{s})$ indicates that the network describes production and decay of $I^1$.

The initial value problem for the deterministic model of~(\ref{eq:outputnet}) is given by
\begin{align}
    \frac{\mathrm{d} x}{\mathrm{d} t}  & =  k_1 (c - x) y - k_2 x, \nonumber\\
		\frac{\mathrm{d} y}{\mathrm{d} t} & = \frac{1}{\mu} \left( 1 - (c - x) y \right), \nonumber \\
		x(0) & = x_0, \nonumber \\
		y(0) & = y_0, \label{eq:outputODEs}
\end{align}
where $x = x(t) \in [0,c] \cap \mathbb{R}_{\ge}$, and $y = y(t) \in \mathbb{R}_{\ge}$, are the concentrations of species $s$, and $I^1$, from~(\ref{eq:outputnet}), respectively, with $x_0, y_0, c \in \mathbb{R}_{\ge}$. We have used the kinetic conservation law $\bar{x}(t) = c - x(t)$, where $\bar{x}(t)$ is the concentration of species $\bar{s}$, and $c < \infty$ is a time-independent conservation constant. Note that the conservation law truncates $x$-state-space. Let us now describe relationships between systems~(\ref{eq:inputODEs}) and~(\ref{eq:outputODEs}), starting with the weak statement: for $c > k_1/k_2$, and for any $\mu \ge 0$, solutions of~(\ref{eq:inputODEs}) and~(\ref{eq:outputODEs}) are the same in the long-time limit $t \to \infty$. More precisely, the $x$-component of the equilibrium point of~(\ref{eq:outputODEs}) is identical to the equilibrium point of~(\ref{eq:inputODEs}), and both are stable. In \textbf{Supplementary Information} (SI) Text, we justify the strong statement: for sufficiently large $c$, and for $\mu \ll 1$, solutions of~(\ref{eq:inputODEs}) and~(\ref{eq:outputODEs}), with the same initial conditions, are approximately the same at each time $t \ge 0$. For these reasons, we call $\mathcal{R}_1^2$ a \emph{drift-corrector network}. 

\subsection{Zero-Drift Network $\mathcal{R}_{1,1}^3$}
Having completed the first two steps, let us focus on the third (and final) step, in which we introduce arbitrary noise into the stochastic model of~(\ref{eq:outputnet}), without influencing the deterministic model~(\ref{eq:outputODEs}). Let us start our consideration by embedding into~(\ref{eq:outputnet}) network $\mathcal{R}_{1,1}^3 = \mathcal{R}_{1,1}^3(s,\bar{s})$, which is given by
\begin{align}
\mathcal{R}_{1,1}^3(s,\bar{s}):  \;  &  & s + \bar{s} & \xrightarrow[]{k_{1,1}}  2 s, \nonumber \\
 &  &  s + \bar{s} & \xrightarrow[]{k_{1,1}}  2 \bar{s}. \label{eq:zerodrift11}
\end{align}
The subscript in $\mathcal{R}_{1,1}^3$ indicates that the underlying reactions have one molecule of $s$, and one of $\bar{s}$, as reactants. The two reactions in~(\ref{eq:zerodrift11}) preserve the conservation law from~(\ref{eq:outputnet}). Furthermore, they fire with the same rates, with the first reaction leading to a unit-production, while the second to a unit-decay, of species $s$. Consequently, embedding $\mathcal{R}_{1,1}^3$ into~(\ref{eq:outputnet}) does not affect the underlying deterministic model~(\ref{eq:outputODEs}), and we call $\mathcal{R}_{1,1}^3$ a \emph{zero-drift network}. However, $\mathcal{R}_{1,1}^3$ does affect the underlying stochastic model~\cite{ZeroDrift1,ZeroDrift2,ZeroDrift3,Janos1}. To illustrate this, let us consider network $\mathcal{R}_{1,1}^3$ in isolation: the reactions from~(\ref{eq:zerodrift11}) fire when $X(t) \in (0,C)$, but not when $X(t) \in \{0,C\}$, so that $\mathcal{R}_{1,1}^3$ in isolation fires until $X(t)$ takes one of the extreme values $\{0,C\}$. Here, $X(t) \in \mathbb{N}_0$, and $C \in \mathbb{N}$, $C < \infty$, are the copy-number of species $s$ appearing in~(\ref{eq:outputnet}) and~(\ref{eq:zerodrift11}) at time $t \ge 0$, and the conservation constant, respectively. \changecolor{Let us note that a possible biologically-relevant realization of network~(\ref{eq:zerodrift11}), aside from DNA strand-displacement mechanism, is a dimer version of the bifunctional histidine kinase/phosphatase reported in~\cite{Tyson}}. 

\begin{figure*}
\centerline{
\hskip 1mm
\includegraphics[width=0.7\columnwidth]{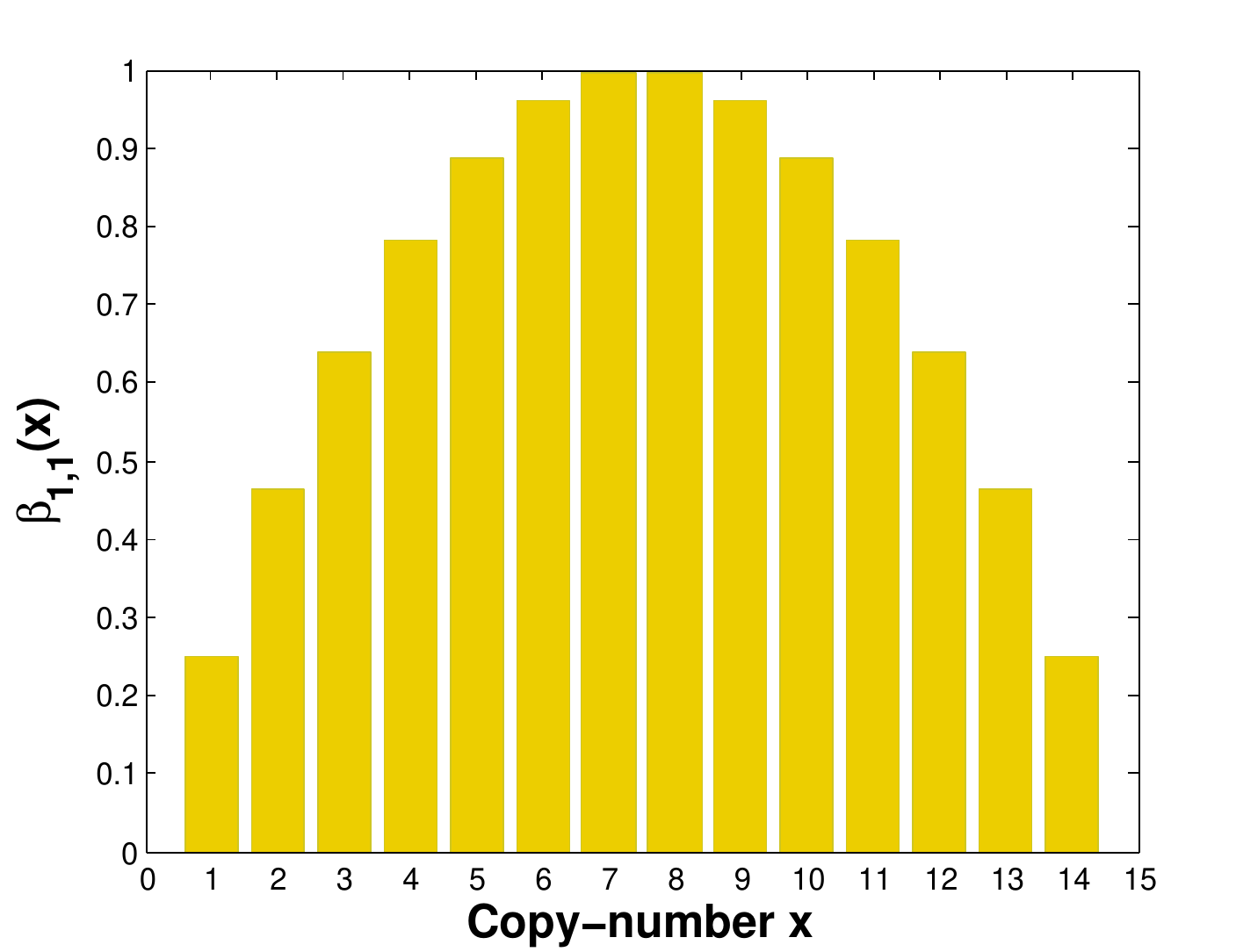}
\hskip 3mm
\includegraphics[width=0.7\columnwidth]{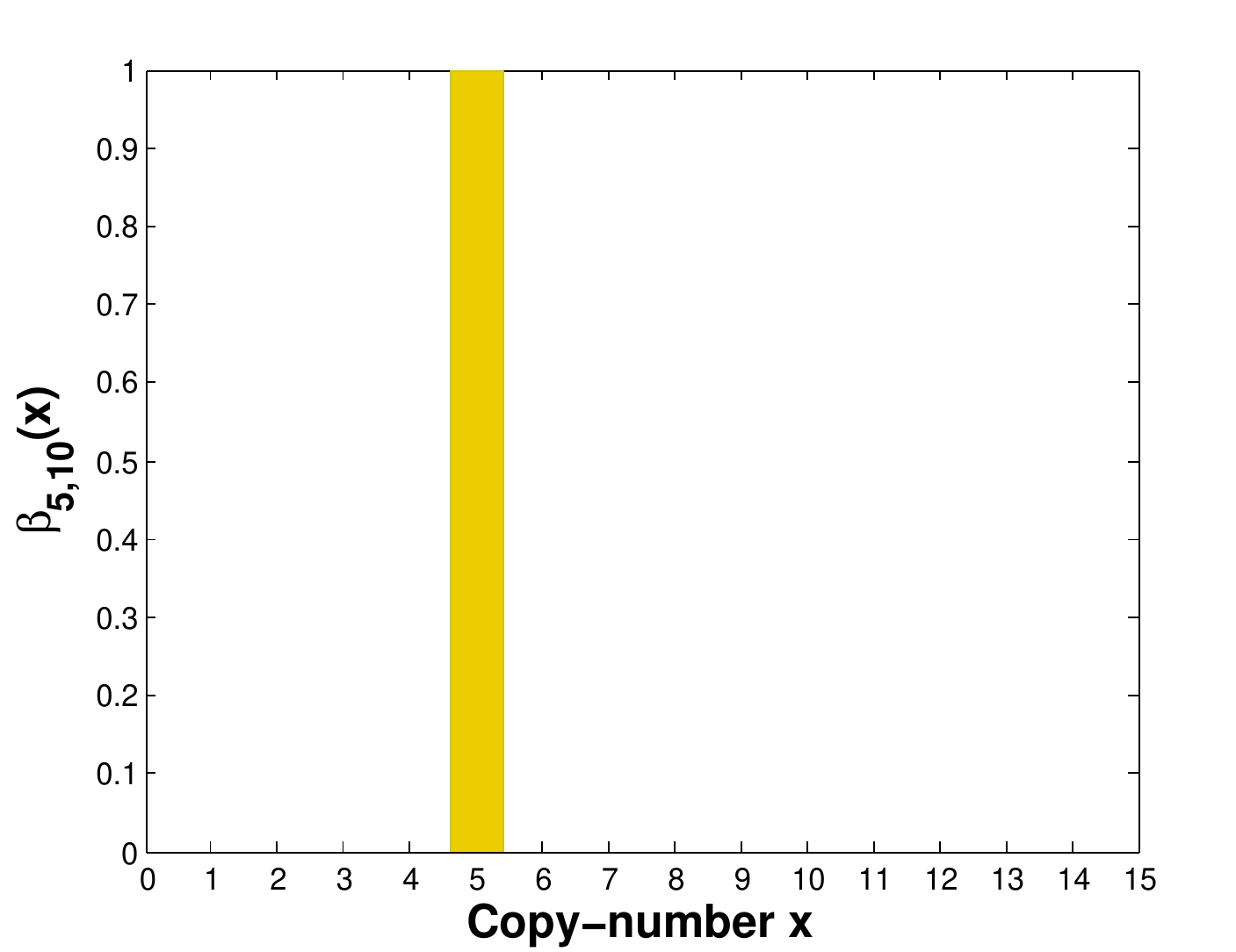}
\hskip 3mm
\includegraphics[width=0.7\columnwidth]{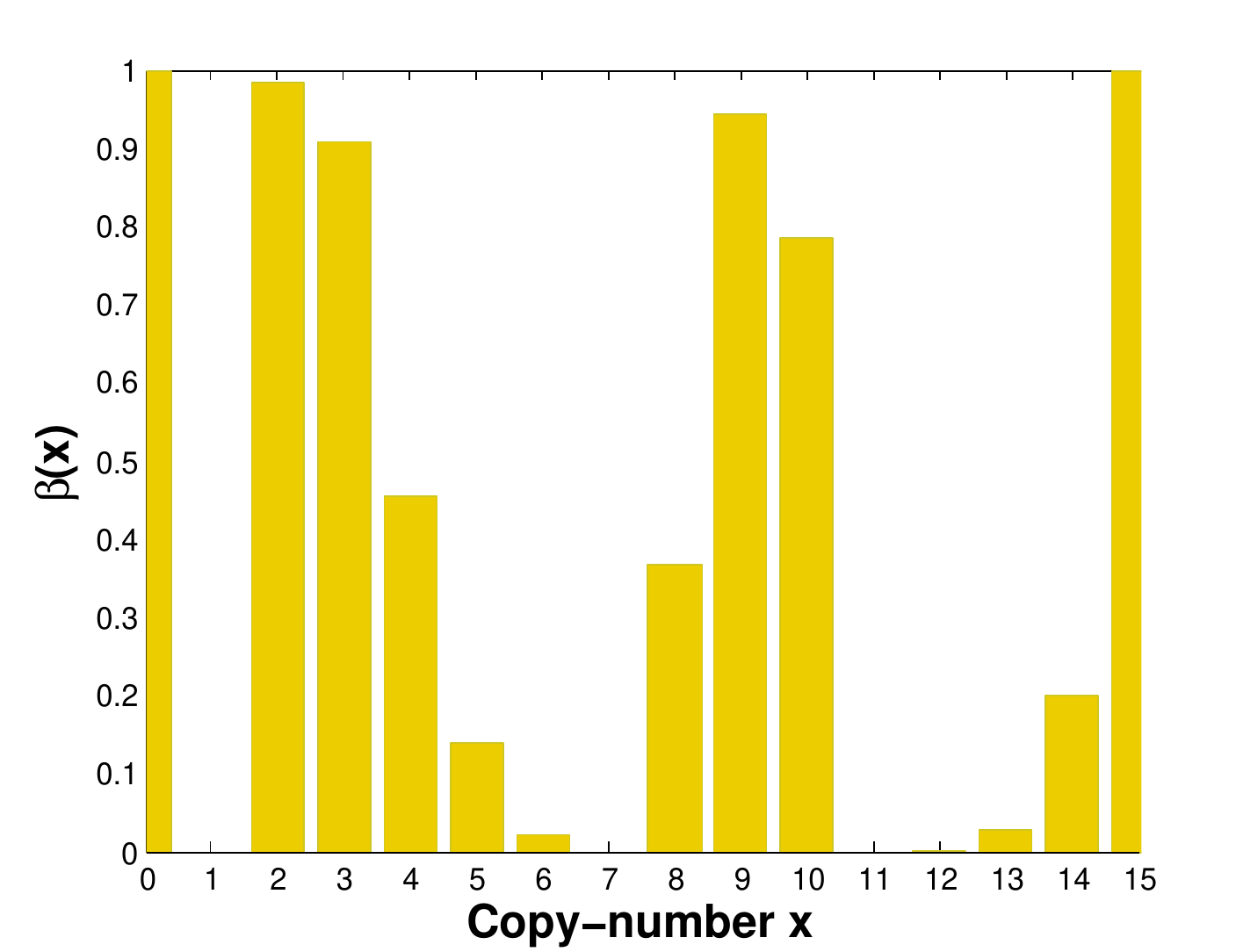}
}
\vskip -4.5cm
\leftline{\hskip -0.7cm (a) \hskip 5.9cm (d) \hskip 6.0cm (g)}
\vskip 4.0cm
\centerline{
\hskip 1mm
\includegraphics[width=0.7\columnwidth]{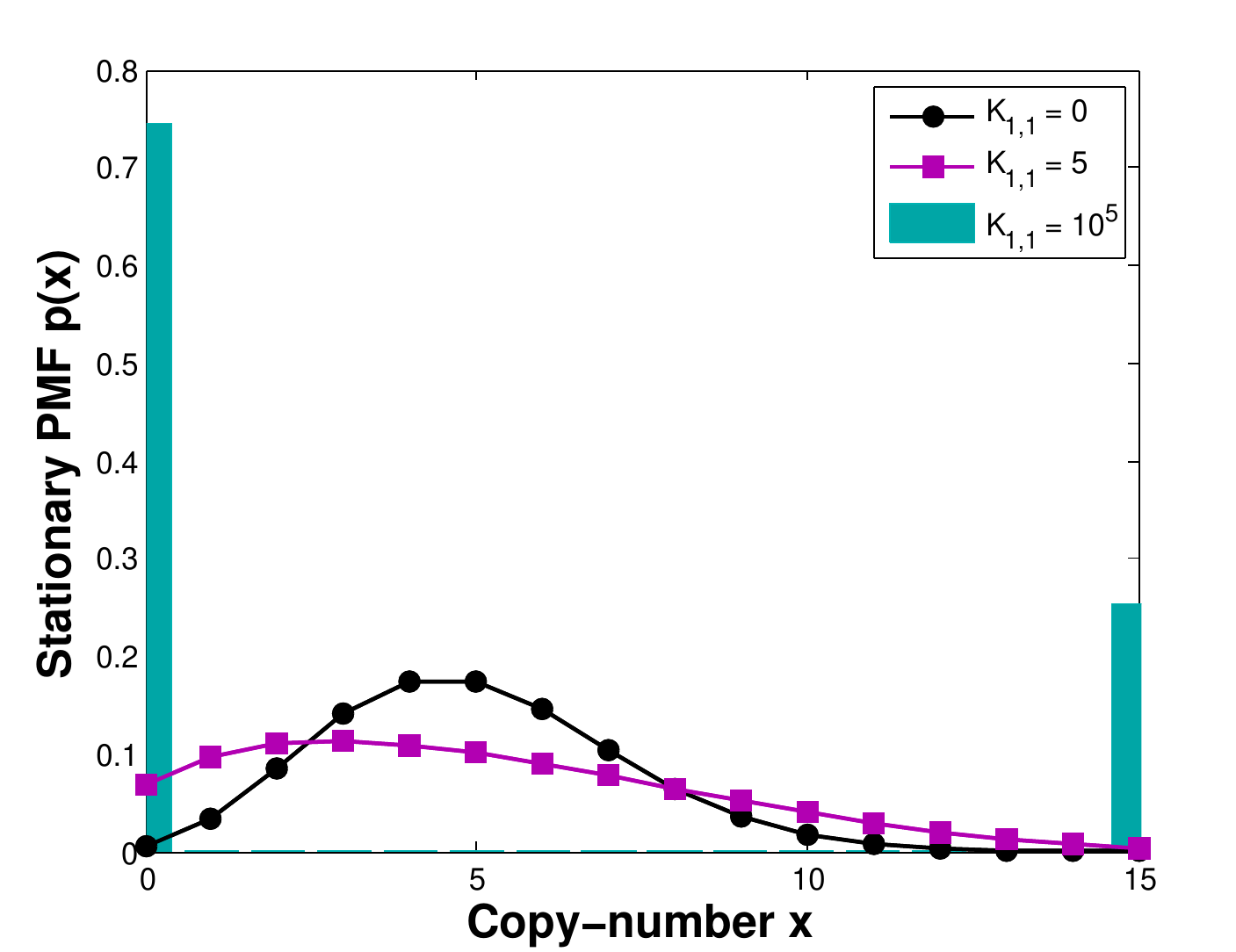}
\hskip 3mm
\includegraphics[width=0.7\columnwidth]{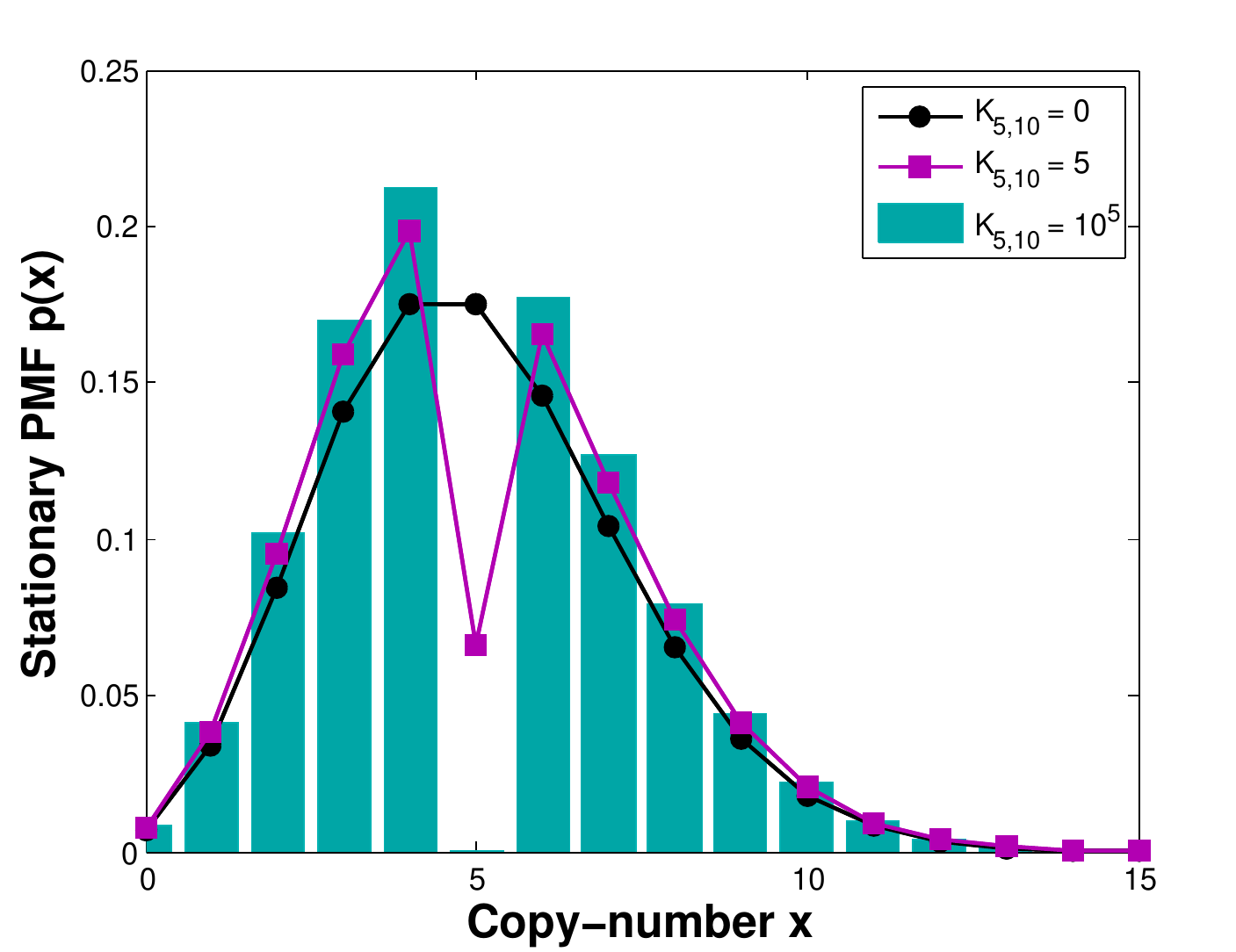}
\hskip 3mm
\includegraphics[width=0.7\columnwidth]{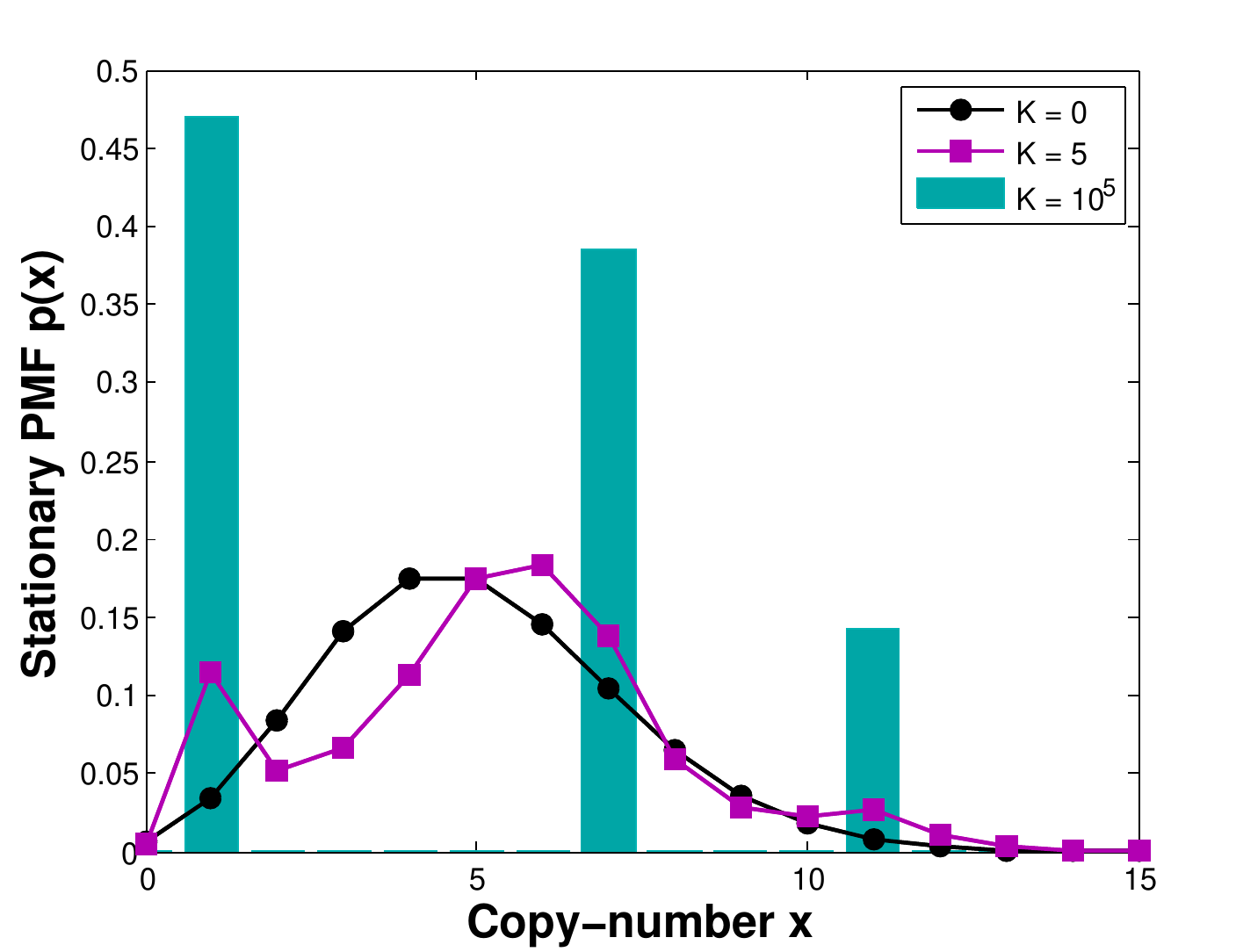}
}
\vskip -4.5cm
\leftline{\hskip -0.7cm (b) \hskip 5.9cm (e) \hskip 6.0cm (h)}
\vskip 4.0cm
\centerline{
\hskip 1mm
\includegraphics[width=0.7\columnwidth]{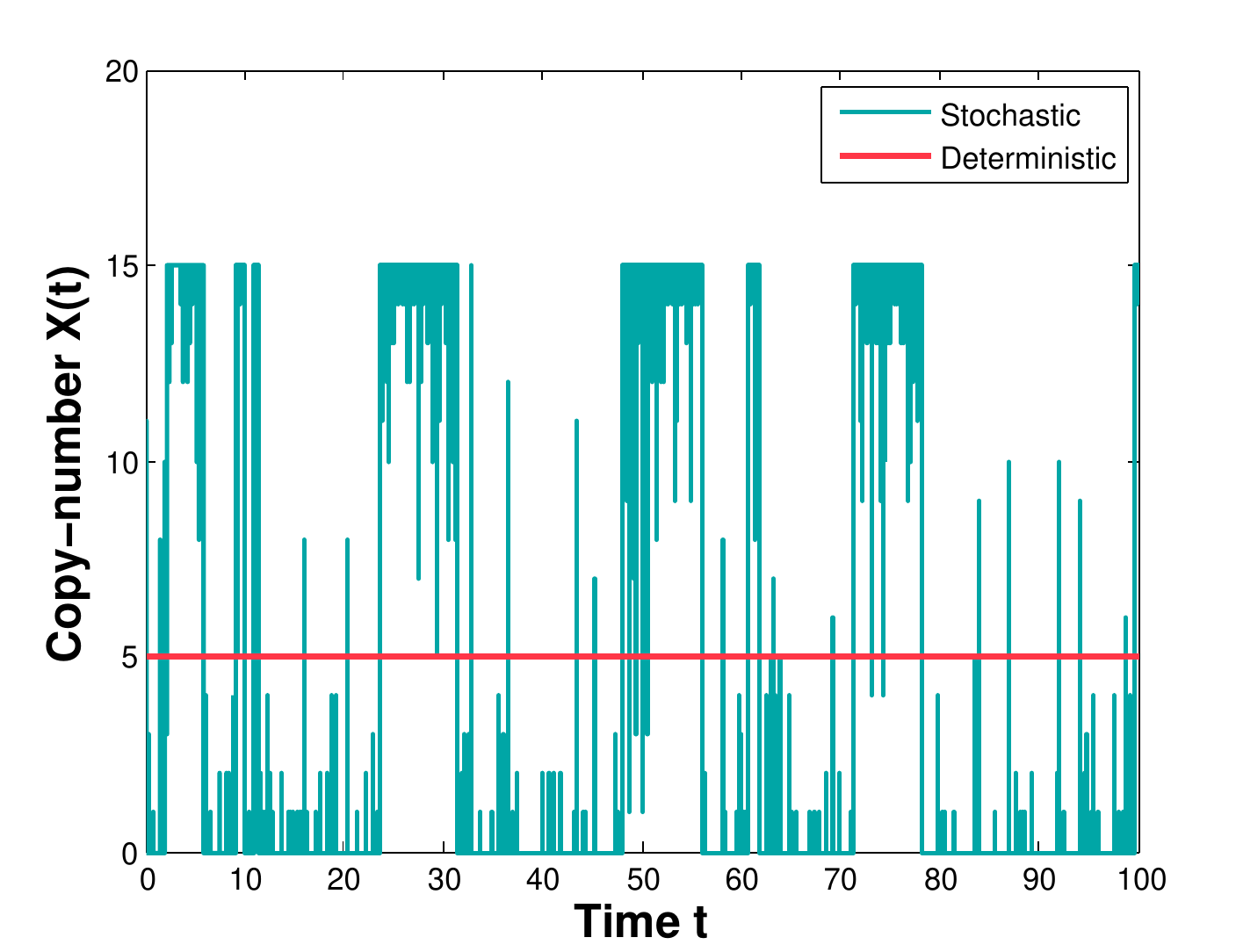}
\hskip 3mm
\includegraphics[width=0.7\columnwidth]{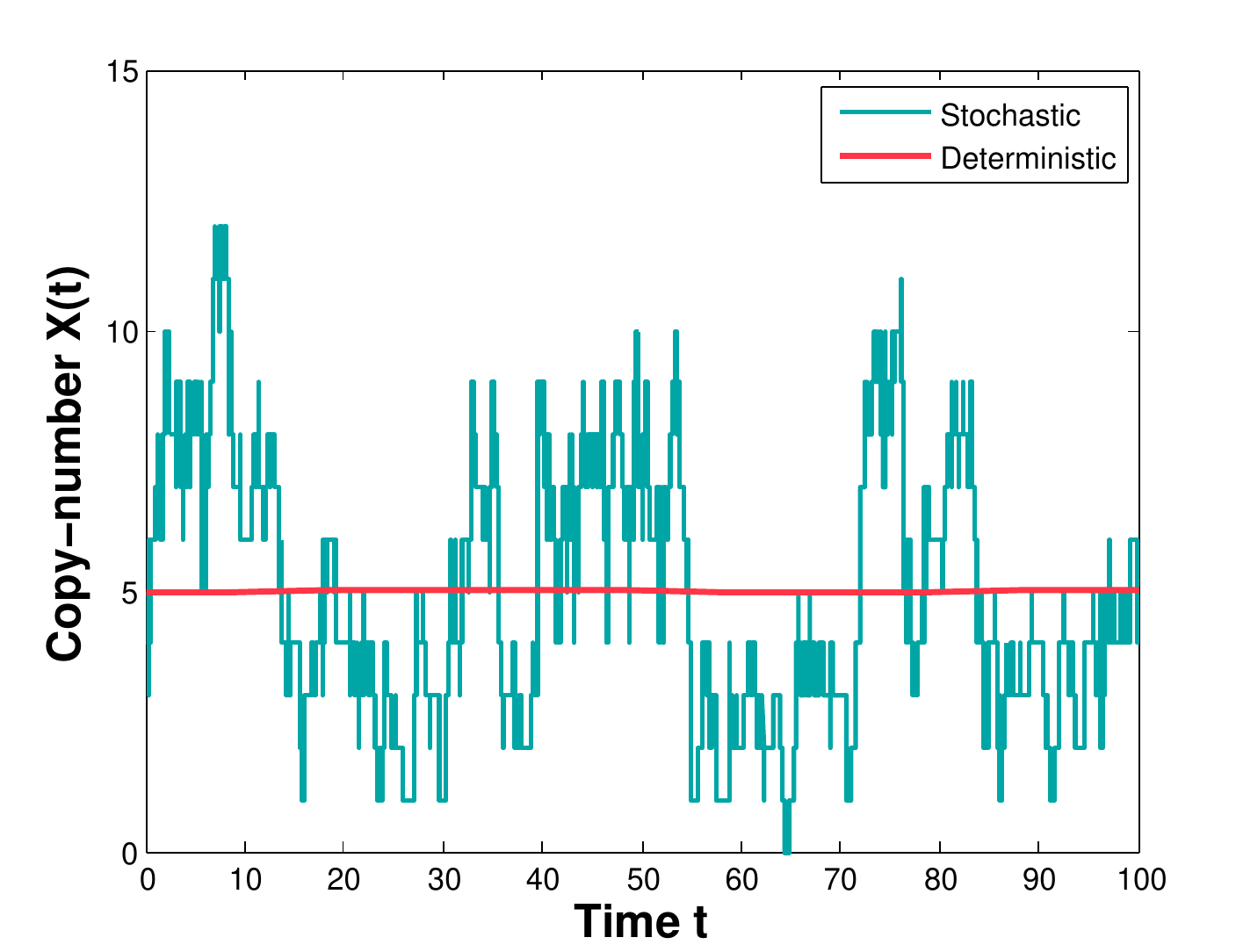}
\hskip 3mm
\includegraphics[width=0.7\columnwidth]{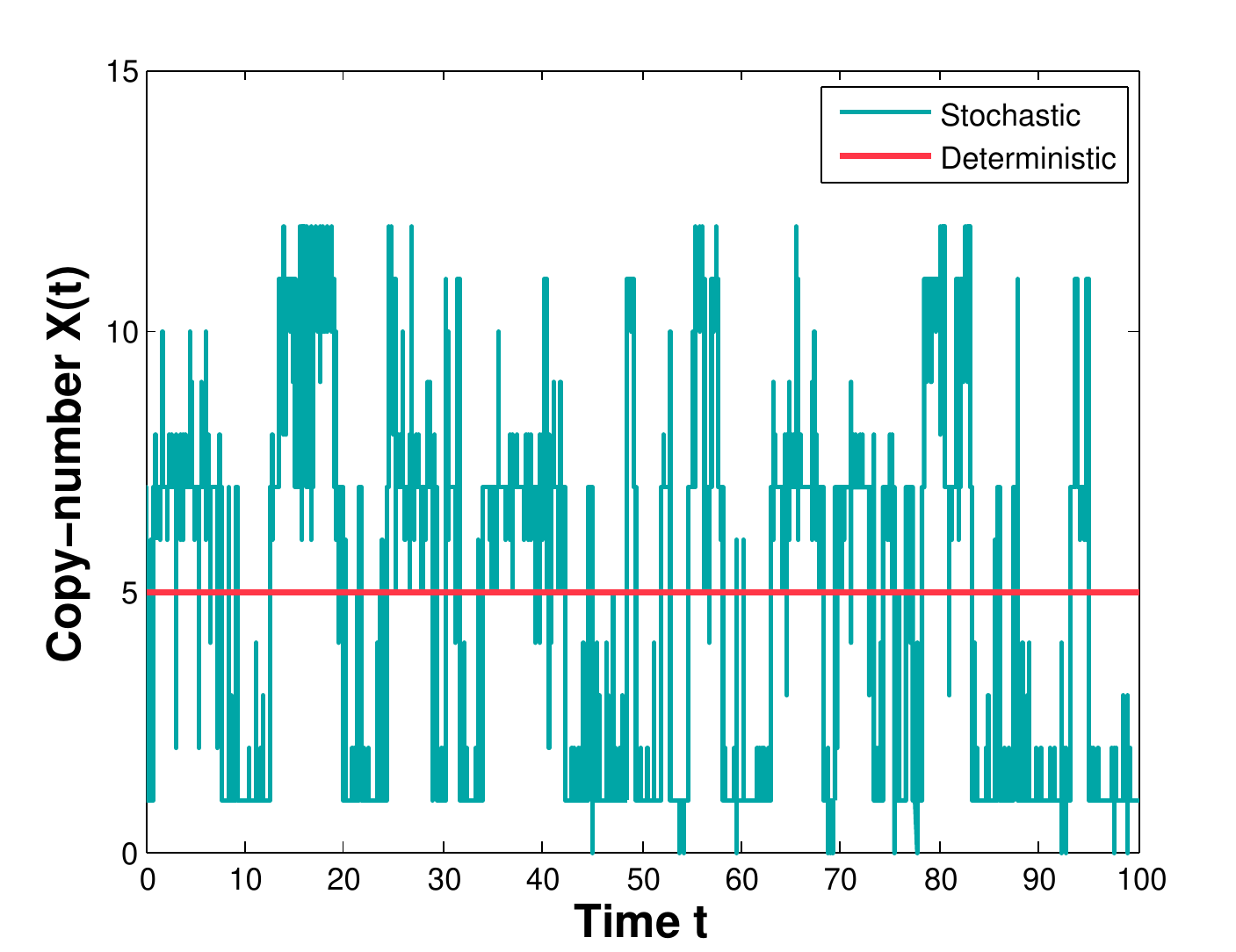}
}
\vskip -4.5cm
\leftline{\hskip -0.7cm (c) \hskip 5.9cm (f) \hskip 6.0cm (i)}
\vskip 4.0cm
\caption{ 
{\it Panels $(a)$, $(d)$ and $(g)$ display propensity functions $\beta_{1,1}(x)$, $\beta_{5,10}(x)$ and $\beta(x)\equiv \beta_{0,15}(x)+ \beta_{2,9}(x) + \beta_{8,5}(x)$ $+ \beta_{12,0}(x)$, respectively. Panels $(b)$, $(e)$ and $(h)$ display the stationary PMF of networks $\hat{\mathcal{R}}^1 \cup \mathcal{R}_1^2 \cup \mathcal{R}_{1,1}^3$, $\hat{\mathcal{R}}^1 \cup \mathcal{R}_1^2 \cup \mathcal{R}_{5,10}^2$ and $\hat{\mathcal{R}}^1 \cup \mathcal{R}_1^2 \cup (\mathcal{R}_{0,15}^3 \cup \mathcal{R}_{2,9}^3 \cup \mathcal{R}_{8,5}^3 \cup \mathcal{R}_{12,0}^3)$, respectively, where $\hat{\mathcal{R}}^1 \cup \mathcal{R}_1^2$ is given by~(\ref{eq:outputnet}), while the rest of the (zero-drift) networks are as given in second step of Algorithm~\ref{tab:thealgorithm}. In $(h)$, $K \equiv K_{0,15} = K_{2,9} = K_{8,5} = K_{12,0}$. Panels $(c)$, $(f)$, and $(i)$ display in blue the sample paths, corresponding to the PMFs shown as the blue histograms in $(b)$, $(e)$ and $(h)$, respectively, and were  obtained by applying the Gillespie algorithm on the underlying networks. Also shown in red are the deterministic trajectories, obtained by numerically solving the corresponding deterministic models. The dimensionless parameters are fixed to: $k_1 = 2.5$, $k_2 = 0.5$, $\mu = 10^{-3}$, $C = 15$, and the state-space for species $I^1$ is bounded in $(b)$, $(e)$ and $(h)$ by $50$. In $(b)$ and $(e)$, the two-species stationary chemical master equation (CME) was numerically solved, while in $(h)$ the boundary zero-drift networks are taken in the asymptotic limits $\mu_{0,15}, \mu_{12,0} \to 0$. The blue and red trajectories from panel $(i)$ were generated with $(\mu_{0,15})^{-1} M_{0,15} = (\mu_{12,0})^{-1} M_{12,0} = 10^{7}$. The trajectories from $(c)$, $(f)$ and $(i)$ were all initiated at the deterministic equilibrium, $X(0) = 5$.}} \label{fig:example1}
\end{figure*}

In SI Text, we derive equation (SI7) which describes the effective behavior of the Markov chain $X(t)$ from network $\hat{\mathcal{R}}^1 \cup \mathcal{R}_1^2 \cup \mathcal{R}_{1,1}^3$ in the limit $\mu \to 0$, and it follows that the effective total propensity function of the network, denoted $\alpha(x)$, satisfies
 \begin{align}
\alpha(x) & \approx \hat{\alpha}(x) + 2 K_{1,1} \beta_{1,1}(x), \, \, \, \, \, \textrm{as } \mu \to 0, \label{eq:effectivepropensity} \\
\hat{\alpha}(x) & = k_1 + k_2 x. \label{eq:totalpropensityinput} 
\end{align}
Function $\hat{\alpha}(x)$ has the form of the total propensity of network~(\ref{eq:inputnet}), and $K_{1,1} \beta_{1,1}(x)$ is the propensity function of reactions in~(\ref{eq:zerodrift11}), with the scaled factors given by
\begin{align}
K_{1,1} & = \left(\frac{C}{2}\right)^2 k_{1,1}, \, \, \, \, \, \, \, \beta_{1,1}(x) = \left(\frac{C}{2} \right)^{-2} x (C-x). \label{eq:auxiliaryfunctions}
\end{align} 
Function $\beta_{1,1}(x)$ is displayed in Figure~\ref{fig:example1}(a), where one can notice its parabolic shape, arising from the underlying conservation law $X(t) + \bar{X}(t) = C$, which holds for all $t \ge 0$, where $\bar{X}(t) \in \mathbb{N}_0$ is the copy-number of $\bar{s}$ at time $t \ge 0$. Comparing~(\ref{eq:effectivepropensity}) and~(\ref{eq:totalpropensityinput}), it follows that, as $\mu \to 0$, the mean interevent time for $X(t)$, from network $\hat{\mathcal{R}}^1 \cup \mathcal{R}_1^2 \cup \mathcal{R}_{1,1}^3$, is lower than that of $\hat{X}(t)$, from network~(\ref{eq:inputnet}), in the regions of the common state-space where $\beta_{1,1}(x) \ne 0$, i.e. for $x \in (0,C)$. Coefficient $K_{1,1}$ controls by how much the interevent time is reduced. Equivalently, $\beta_{1,1}(x)$, and $K_{1,1}$, determine the support, and magnitude, respectively, of the state-dependent intrinsic noise which network~(\ref{eq:zerodrift11}) introduces into the dynamics of network~(\ref{eq:outputnet}). 

To study this further, in SI Text we derive the following two equations (given as (SI9), and (SI13), respectively)
\begin{align}
\lim_{K_{1,1} \to 0} p(x) & \approx
  \begin{cases}
  \frac{1}{x!} \left( \frac{k_1}{k_2}\right)^{x} \exp \left(- \frac{k_1}{k_2}\right), & \textrm{if } x \in [0,C], \\
   0,   & \text{otherwise},  \label{eq:inputstationaryPMF}
  \end{cases}
  \\
\lim_{K_{1,1} \to \infty} p(x) & \approx
  \begin{cases}
   1 - \frac{1}{C} \frac{k_1}{k_2}, & \text{if } x = 0, \\
   \frac{1}{C} \frac{k_1}{k_2}, & \text{if } x = C, \\
	 0, & \text{otherwise},  \label{eq:f02}
  \end{cases}
\end{align}
where $p(x)$ is the stationary probability mass function (PMF) corresponding to network $\hat{\mathcal{R}}^1 \cup \mathcal{R}_1^2 \cup \mathcal{R}_{1,1}^3$ in the limit $\mu \to 0$, i.e. the probability that there are $x$ molecules of species $s$ as $\mu \to 0$ in the long-time limit $t \to \infty$. Let us interpret analytical results~(\ref{eq:inputstationaryPMF}) and~(\ref{eq:f02}), and compare them with the numerically obtained counterparts. In Figure~\ref{fig:example1}(b), we display numerically obtained stationary $x$-marginal PMFs for different values of $K_{1,1}$, with the rest of the (dimensionless) parameters fixed to $k_1 = 2.5$, $k_2 = 0.5$, $\mu = 10^{-3}$, and $C = 15$. It can be seen that, for $K_{1,1} = 0$, i.e. when the zero-drift network $\mathcal{R}_{1,1}^3$ does not fire, the PMF matches that of network~(\ref{eq:inputnet}), i.e. it is a Poissonian, as predicted by~(\ref{eq:inputstationaryPMF}). Let us note that the matching of the PMFs of networks~(\ref{eq:inputnet}) and $\hat{\mathcal{R}}^1 \cup \mathcal{R}_1^2 \cup \mathcal{R}_{1,1}^3$ relies on choosing sufficiently large rate coefficients $1/\mu$ in the drift-corrector network $\mathcal{R}_{1}^2$. When $K_{1,1} = 5$, the PMF appears closer to a uniform distribution, than does the PMF when $K_{1,1} = 0$. Finally, for the larger value $K_{1,1} = 10^5$, i.e. when zero-drift network $\mathcal{R}_{1,1}^3$ fires much faster than network $\hat{\mathcal{R}}^1$, the PMF redistributes across the domain, accumulating at the boundary, and becoming bimodal. This is in qualitative agreement with~(\ref{eq:effectivepropensity}), and in quantitative agreement with~(\ref{eq:f02}), which predicts $p(0) \approx 0.7$ and $p(15) \approx 0.3$. In Figure~\ref{fig:example1}(c), a representative sample path is shown, obtained by applying the Gillespie algorithm on network $\hat{\mathcal{R}}^1 \cup \mathcal{R}_1^2 \cup \mathcal{R}_{1,1}^3$, when $K_{1,1} = 10^5$. Also shown is a trajectory obtained by numerically solving the deterministic model~(\ref{eq:outputODEs}). Consistent with Figure~\ref{fig:example1}(b), the sample path switches between the boundary of the state-space, with a bias towards the left boundary point $x = 0$. This is in contrast to the deterministic trajectories, which are globally attracted to the equilibrium point $x = 5$.   

\subsection{General Zero-Drift Networks $\mathcal{R}_{n,\bar{n}}^3$}
Zero-drift network $\mathcal{R}_{1,1}^3(s,\bar{s})$, given by~(\ref{eq:zerodrift11}), involves a single molecule of $s$ and $\bar{s}$ as reactants, and adds the noise at $x \in [1,C-1]$, i.e. in the interior of the state-space. Similar networks may be used to add the noise at any point in the state-space, without influencing the deterministic dynamics. In particular, in~(\ref{eq:zerodriftinterior}) and~(\ref{eq:zerodriftleftboundary}), we present general zero-drift networks $\mathcal{R}_{n,\bar{n}}^3(s,\bar{s})$, which involve $n$ molecules of $s$, and $\bar{n}$ of $\bar{s}$, as reactants, and add the noise at $x \in [n,C-\bar{n}]$, where $n,\bar{n} \in \mathbb{N}_0$, and $(n + \bar{n}) \le C$ (see also SI Text). Embedding a union of such networks, $\cup_{(n,\bar{n})} \mathcal{R}_{n,\bar{n}}^3(s,\bar{s})$, into~(\ref{eq:outputnet}), we arrive at the result similar to~(\ref{eq:effectivepropensity}), with $K_{1,1} \beta_{1,1}(x)$ replaced by the linear combination $\sum_{(n, \bar{n})} K_{n,\bar{n}} \beta_{n,\bar{n}}(x)$. The scaled rate coefficient $K_{n,\bar{n}}$, and function $\beta_{n,\bar{n}}(x)$, are given as (S14), and (S15), respectively, in  SI Text, where we also justify that an arbitrary nonnegative function, with compact support, may be approximated by a suitable sum $\sum_{(n, \bar{n})} K_{n,\bar{n}} \beta_{n,\bar{n}}(x)$. To illustrate general zero-drift networks, let us start with embedding into network~(\ref{eq:outputnet}) zero-drift network $\mathcal{R}_{5,10}^3(s,\bar{s})$, satisfying~(\ref{eq:zerodriftinterior}) with $n = 5$ and $\bar{n} = 10$. In Figure~\ref{fig:example1}(d), we show propensity function $\beta_{5,10}(x)$, which is nonzero only at $x = 5$. In~(e), we show the numerically approximated stationary $x$-marginal PMFs underlying network $\hat{\mathcal{R}}^1 \cup \mathcal{R}_1^2 \cup \mathcal{R}_{5,10}^3$ for different values of $K_{5,10}$, with the rest of the parameters as in Figure~\ref{fig:example1}(b). One can notice that, under the action of network $\mathcal{R}_{5,10}^3$, the PMF is gradually decreased to nearly zero at $x = 5$ (the deterministic equilibrium), and becomes bimodal, with the two noise-induced maxima at $x = 4$ and $x = 6$. In~(f), we show a corresponding representative sample path.

In general, noise-induced multimodality may be achieved by a suitable combination of zero-drift networks. For example, let us synthetize noise such that the stationary PMF is trimodal, and nearly zero everywhere, except at $x \in \{1,7,11\}$. Such a task may always be achieved by a suitable combination of the basis zero-drift networks, i.e. those zero-networks that induce noise only at a single point in the state-space (e.g. subnetwork $\mathcal{R}_{5,10}^3$ with propensity function shown in Figure~\ref{fig:example1}(d), see also SI Text). In the present case, one could construct the thirteen basis zero-drift networks which add large enough noise at $x \in [0,15] \setminus \{1,7,11\}$.  Here, for simplicity, we achieve the task with only four zero-drift networks. In Figures~\ref{fig:example1}(g)--(i), we consider network $\hat{\mathcal{R}}^1 \cup \mathcal{R}_1^2 \cup (\mathcal{R}_{0,15}^3 \cup \mathcal{R}_{2,9}^3 \cup \mathcal{R}_{8,5}^3 \cup \mathcal{R}_{12,0}^3)$. We denote $\beta(x) \equiv \beta_{0,15}(x) + \beta_{2,9}(x) + \beta_{8,5}(x) + \beta_{12,0}(x)$, and, for simplicity, take $K \equiv K_{0,15} = K_{2,9} = K_{8,5} = K_{12,0}$. The resultant propensity function $\beta(x)$ is shown in~(g), while in~(h) it can be seen that the PMF becomes trimodal for sufficiently large $K$, with the maxima at $x = \{1,7,11\}$. This is consistent with the corresponding representative sample path shown in blue in panel (i), which display tristability. Let us note that, while the stochastic dynamics display multistability in (c), (f) and (i), the corresponding deterministic dynamics, also shown in the plots, remain monostable. 

\section{A Two-species Exotic System} \label{sec:example2}
\begin{figure*}
\centerline{
\hskip 1mm
\includegraphics[width=0.7\columnwidth]{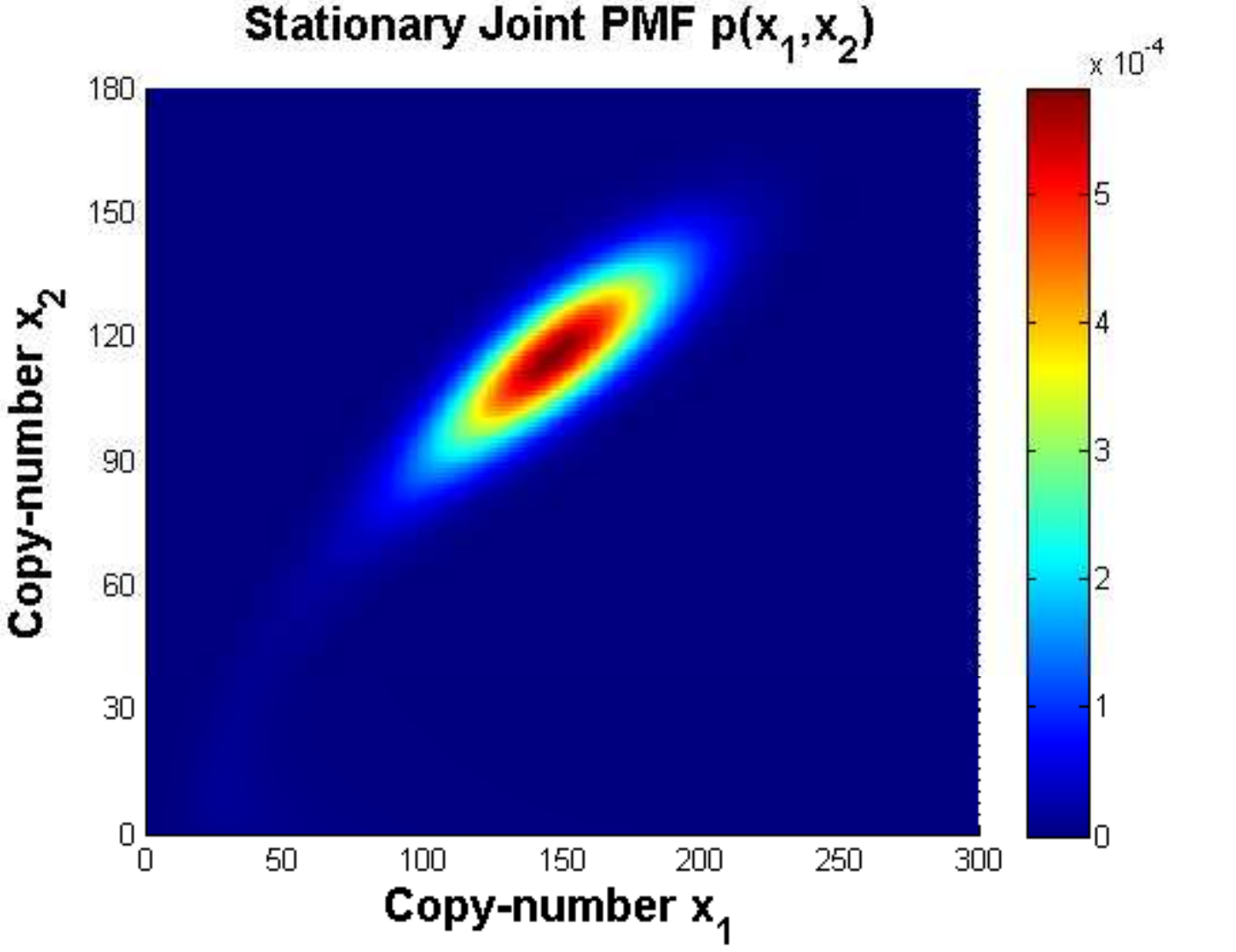}
\hskip 3mm
\includegraphics[width=0.7\columnwidth]{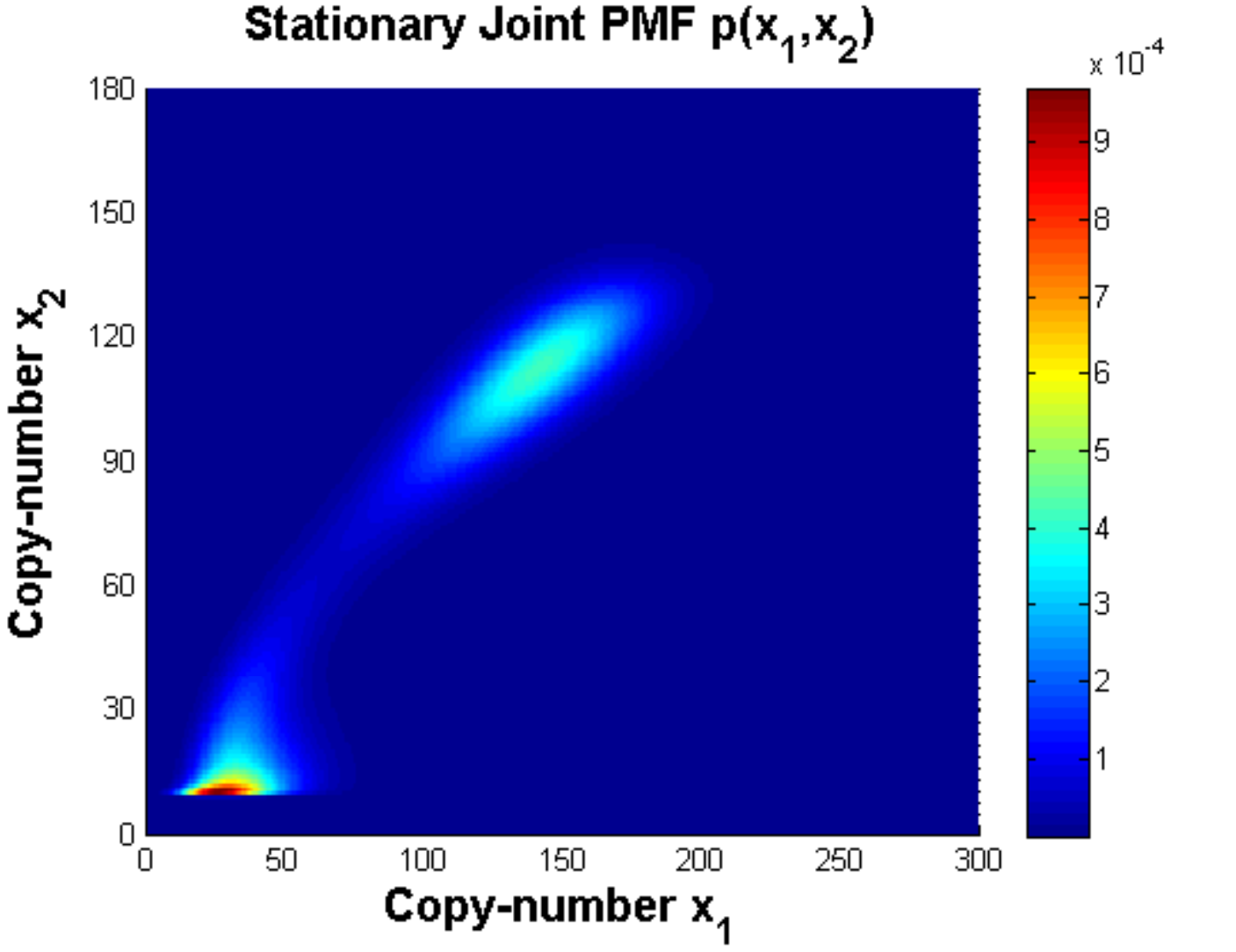}
\hskip 3mm
\includegraphics[width=0.7\columnwidth]{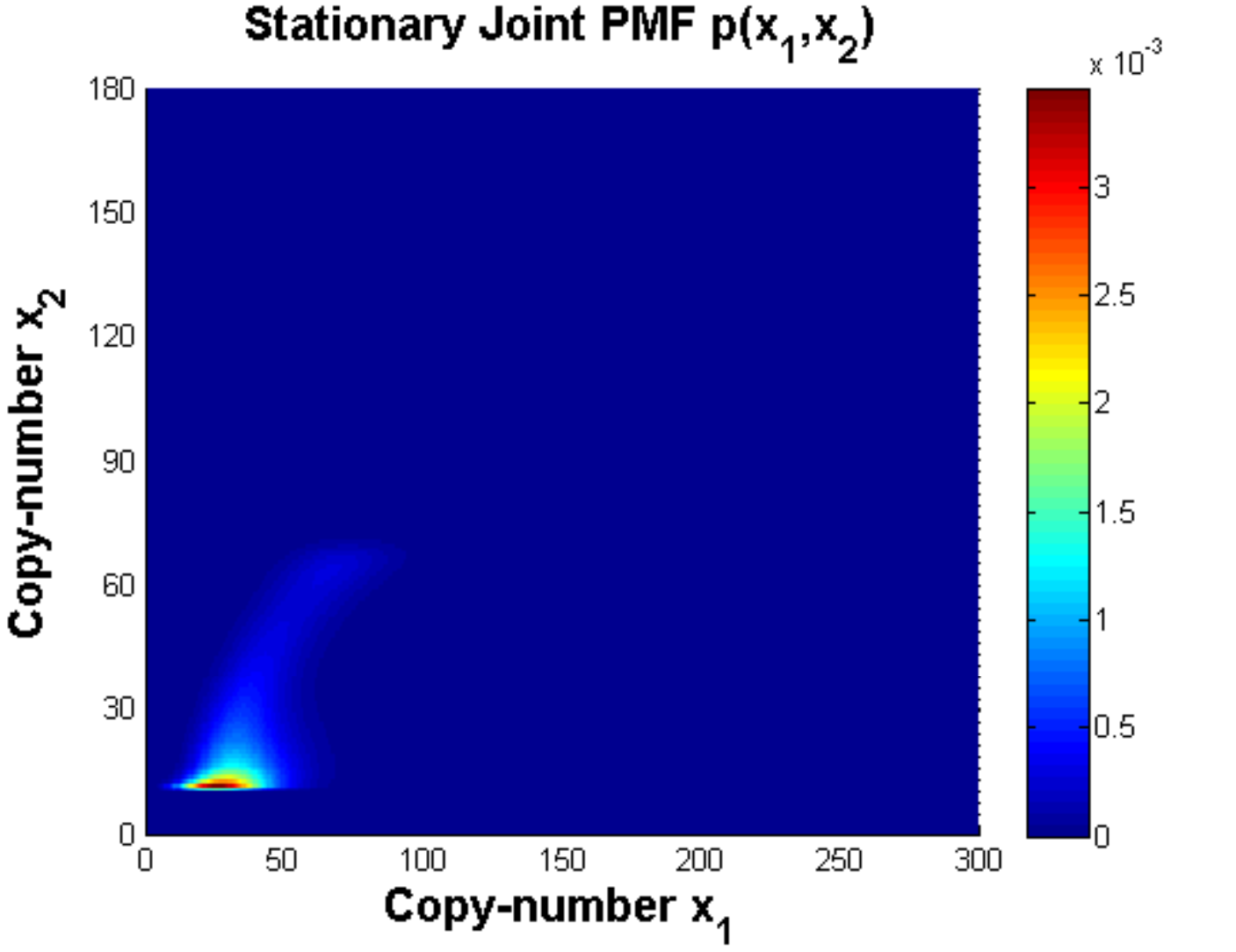}
}
\vskip -4.5cm
\leftline{\hskip -0.7cm (a) \hskip 5.9cm (d) \hskip 6.0cm (g)}
\vskip 4.0cm
\centerline{
\hskip 1mm
\includegraphics[width=0.7\columnwidth]{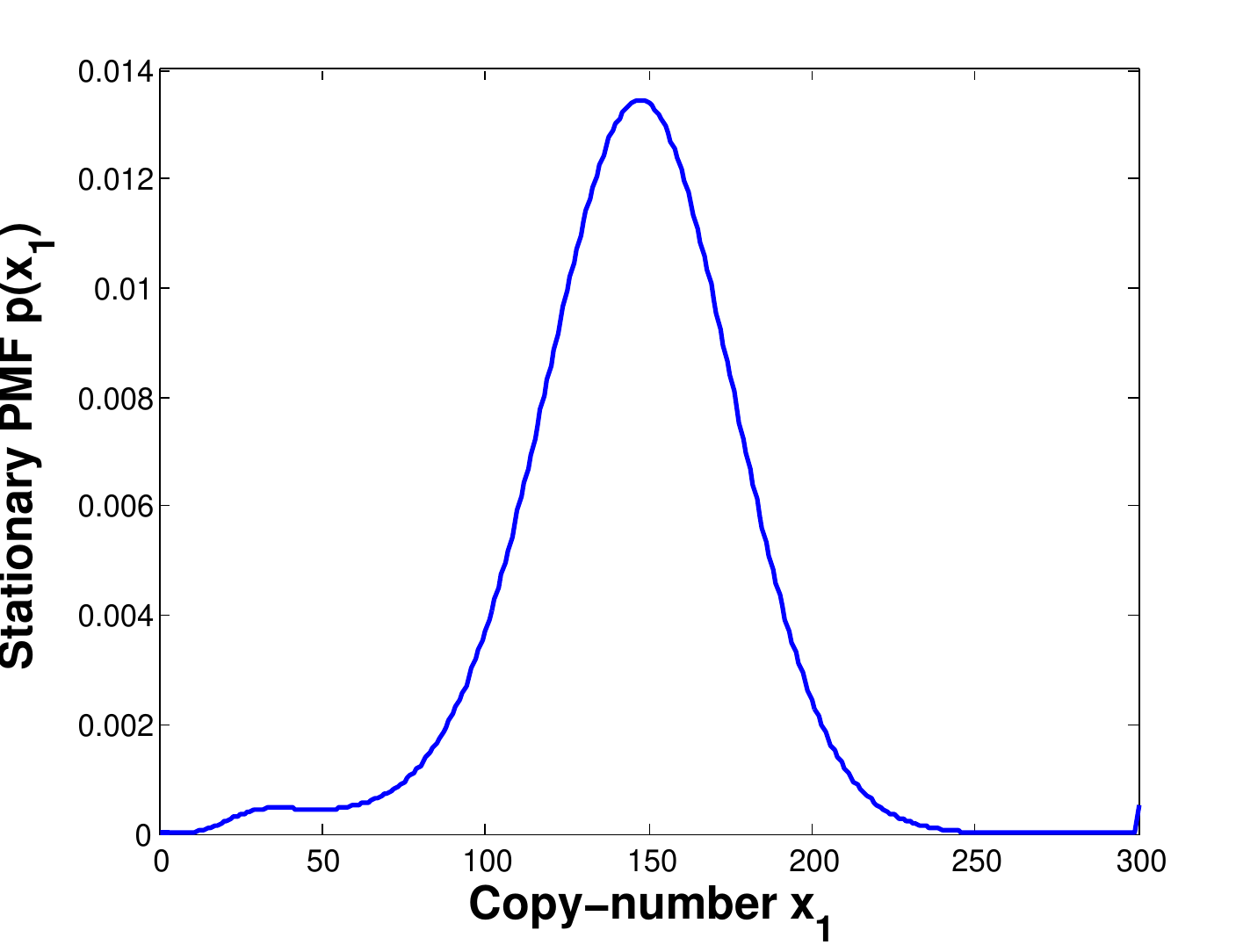}
\hskip 3mm
\includegraphics[width=0.7\columnwidth]{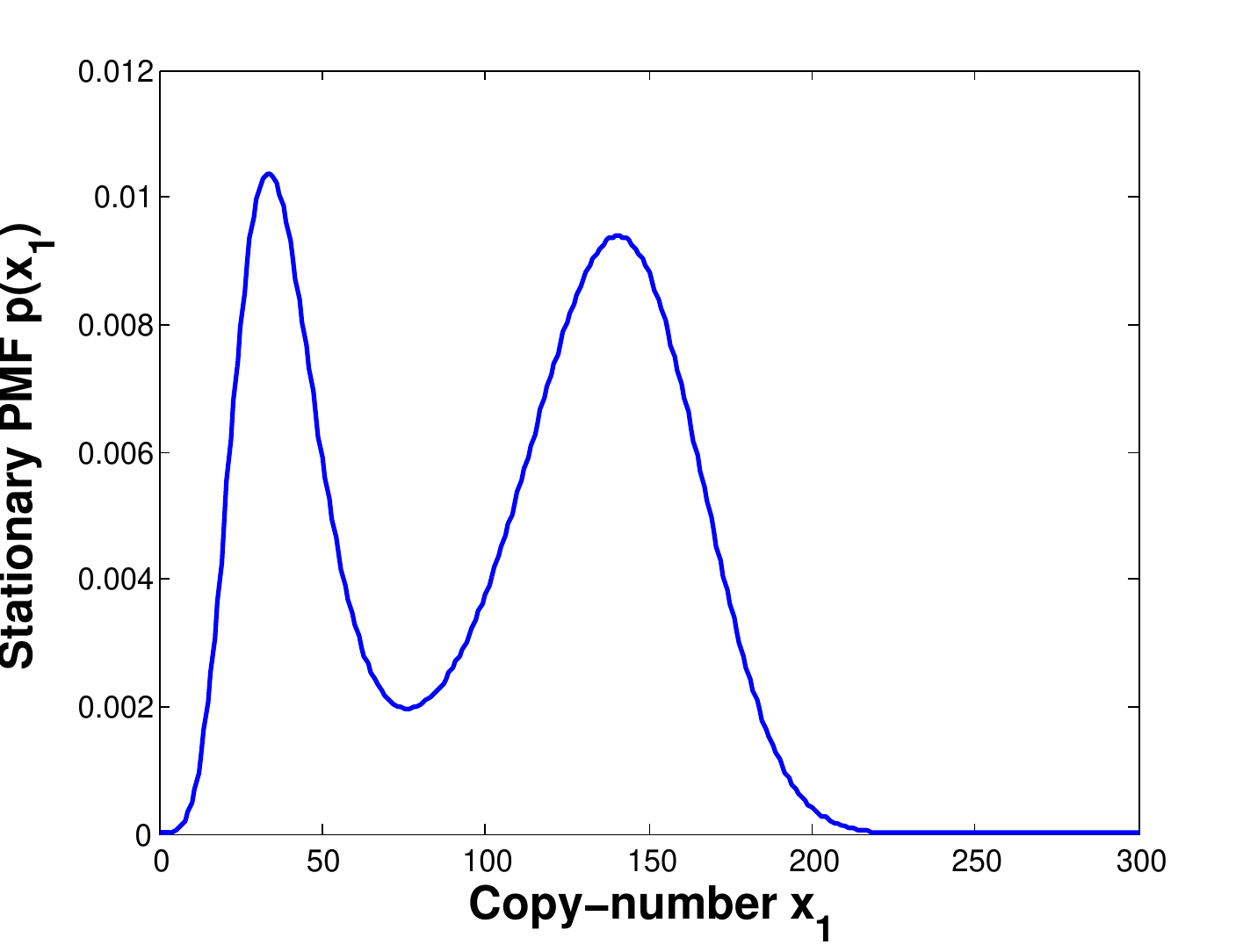}
\hskip 3mm
\includegraphics[width=0.7\columnwidth]{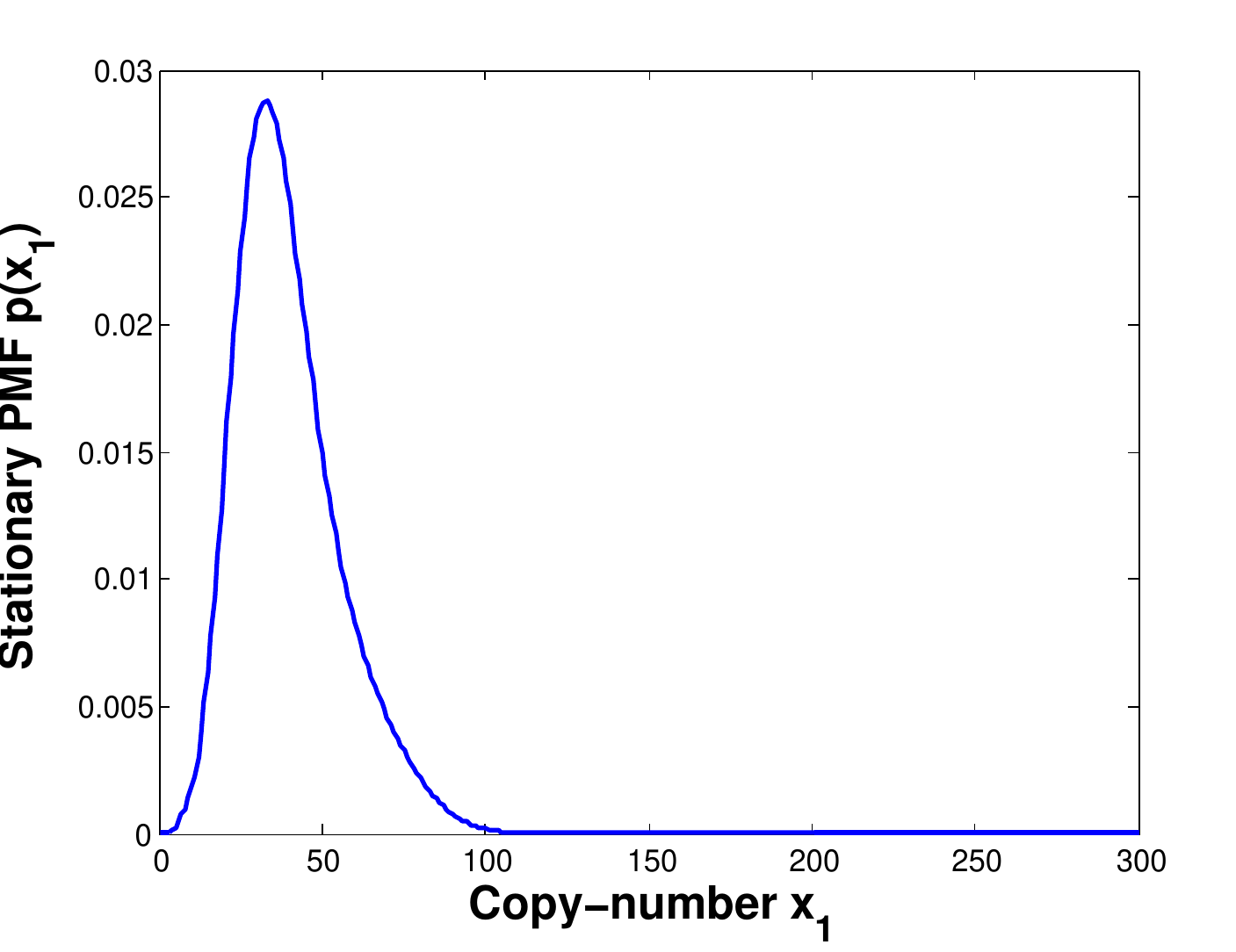}
}
\vskip -4.5cm
\leftline{\hskip -0.7cm (b) \hskip 5.9cm (e) \hskip 6.0cm (h)}
\vskip 4.0cm
\centerline{
\hskip 1mm
\includegraphics[width=0.7\columnwidth]{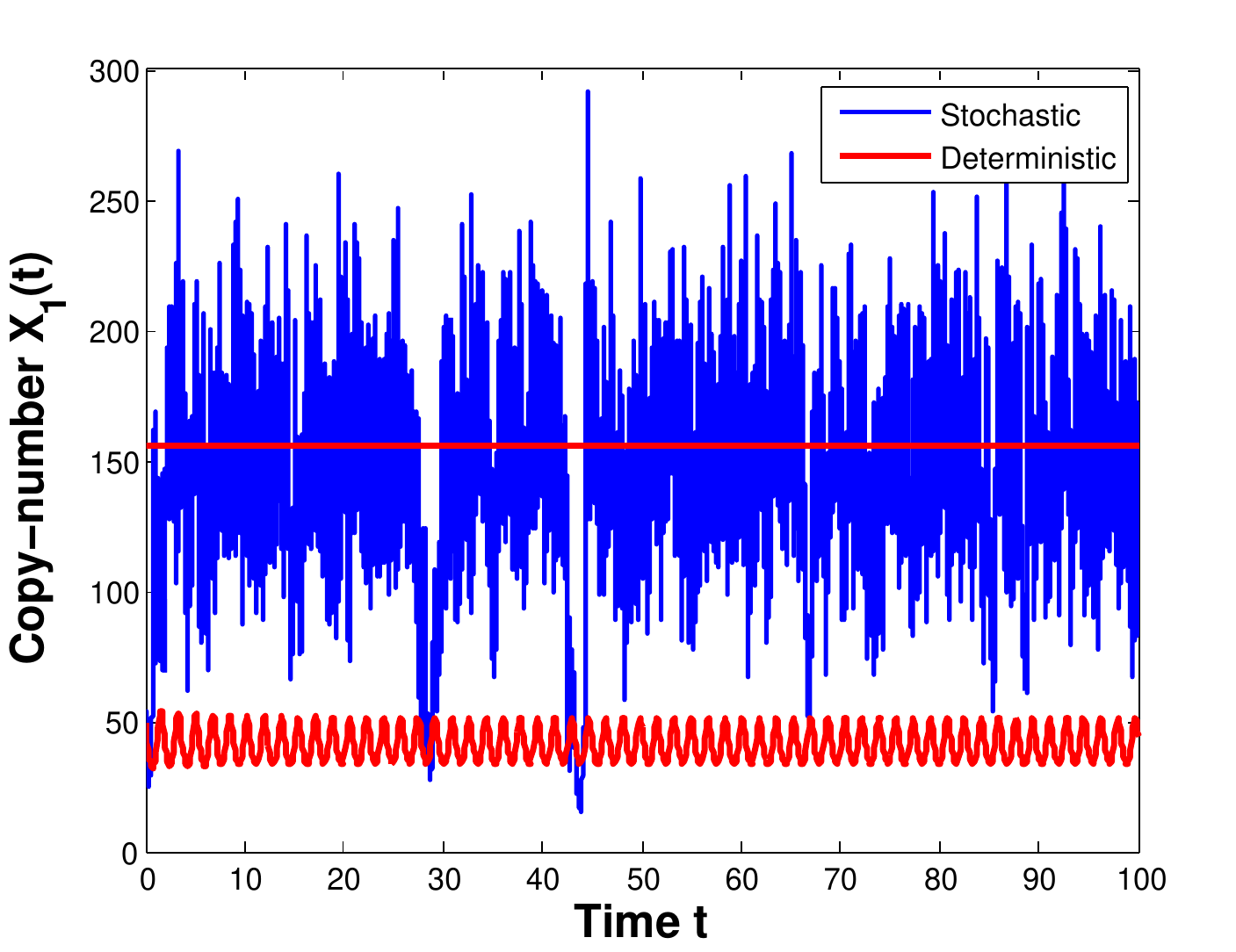}
\hskip 3mm
\includegraphics[width=0.7\columnwidth]{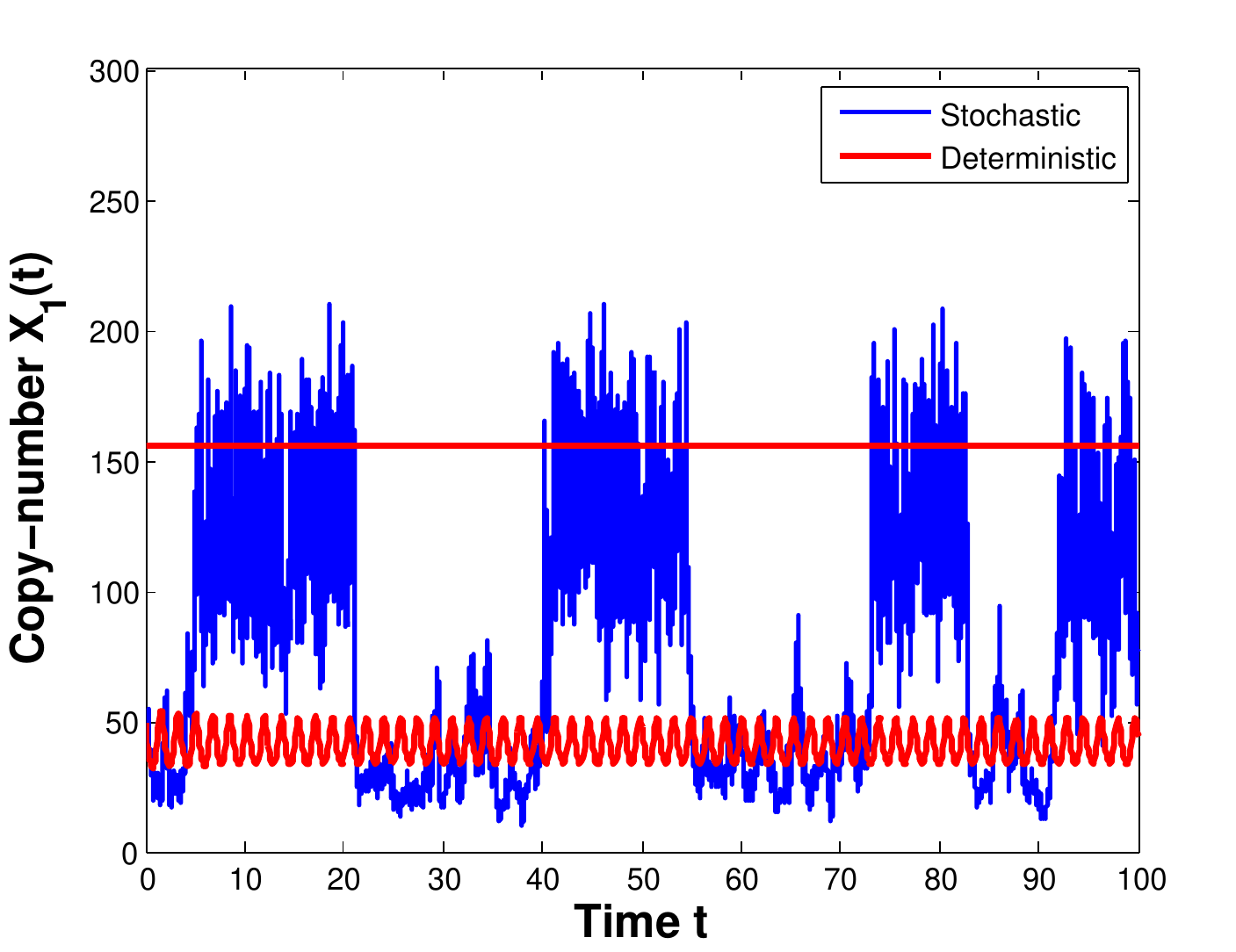}
\hskip 3mm
\includegraphics[width=0.7\columnwidth]{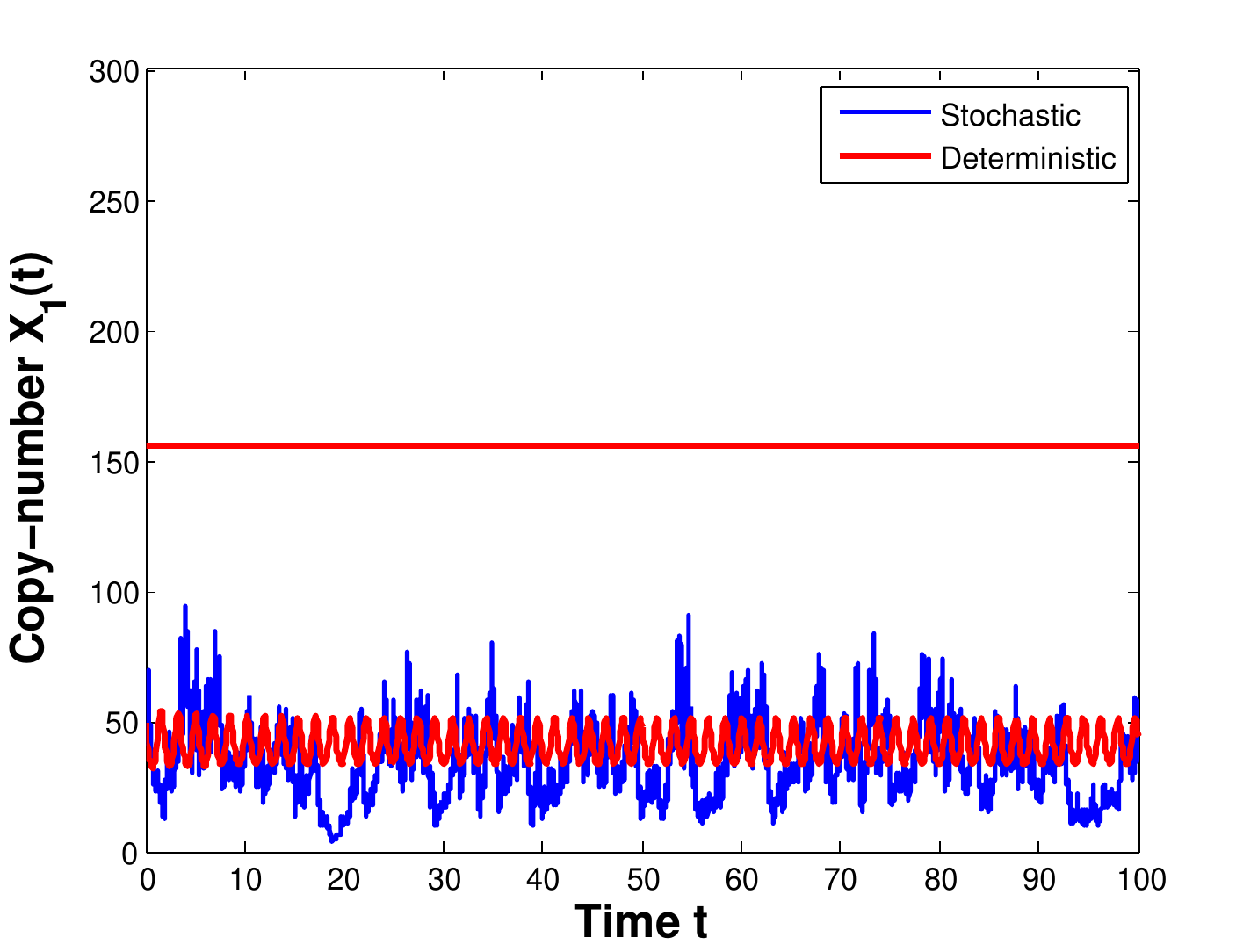}
}
\vskip -4.5cm
\leftline{\hskip -0.7cm (c) \hskip 5.9cm (f) \hskip 6.0cm (i)}
\vskip 4.0cm
\caption{ 
{\it Panel $(a)$ displays the joint stationary PMF of network~(\ref{eq:inputnet2}), while $(d)$ and $(g)$ display the stationary PMFs of network (SI18) from SI Text for $(K_{0,C_2-10},K_{30,0}) = (10^{18},2 \times 10^8)$ and $(K_{0,C_2-10},K_{30,0}) = (10^{18},10^{18})$, respectively, with the rest of the parameters being the same. Panels $(b)$, $(e)$ and $(h)$ display the $x_1$-marginal PMFs corresponding to $(a)$, $(b)$ and $(c)$, respectively. Panels $(c)$, $(f)$ and $(i)$ display in blue the sample paths, corresponding to the PMFs shown in $(b)$, $(e)$ and $(h)$, respectively, and were obtained by applying the Gillespie algorithm on the underlying networks. Also shown in red are two deterministic trajectories, one initiated near the equilibrium point, while the other near the limit cycle, obtained by numerically solving equation $(SI17)$ from SI Text. The dimensionless parameters are fixed to: $k_1 = 4$, $k_2 = 1.408$, $k_3 = 0.0518$, $k_4 = 0.164$, $k_5 = 3.1 \times 10^{-3}$, $k_6 = 4.8 \times 10^{-3}$, $k_7 = 4$, $k_8 = 8$, $k_9 = 0.16$, $k_{10} =  0.104$, $k_{11} = 2.1 \times 10^{-3}$. In $(a)$--$(b)$, $(d)$--$(e)$ and $(g)$--$(h)$, the stationary chemical master equation (CME) is numerically solved, with the state-space is truncated to $(x_1,x_2) \in [0,C_1] \times [0,C_2]$, where $C_1 = 300$, $C_2 = 180$, and $\mu, \mu_{0,C_2 - 10}, \mu_{30,0} \to 0$. The blue sample paths from panels $(f)$ and $(i)$ were generated with $(\mu^{-1}, (\mu_{0,C_2 - 10})^{-1} M_{0,C_2 - 10}, (\mu_{30,0})^{-1} M_{30,0}) =  (10^{3}, 10^{20}, 2 \times 10^{10})$ and $(\mu^{-1}, (\mu_{0,C_2 - 10})^{-1} M_{0,C_2 - 10}, (\mu_{30,0})^{-1} M_{30,0}) =  (10^{3}, 10^{20}, 10^{20})$, respectively. The blue trajectories from $(c)$, $(f)$ and $(i)$ were all initiated near the deterministic limit cycle.}} \label{fig:example2}
\end{figure*}

Consider the two-species network $\tilde{\mathcal{R}}(s_1,s_2)$, given by
\begin{align}
\tilde{\mathcal{R}}(s_1,s_2): & & \varnothing  &
\xrightarrow[]{ k_1 } s_1,  
& 
\varnothing  &\xrightarrow[]{ k_7 } s_2,  \nonumber \\
&  & s_1  &
\xrightarrow[]{ k_2 } 2 s_1, 
& 
s_2  &\xrightarrow[]{ k_8 } \varnothing,  \nonumber \\
& & 2 s_1  &\xrightarrow[]{ k_3 } 3 s_1,  & s_1 + s_2  &\xrightarrow[]{ k_9 } s_1 + 2 s_2, \nonumber \\
&  & s_1 + s_2  &\xrightarrow[]{ k_4 } s_2, 
&
2 s_2  &\xrightarrow[]{ k_{10} } 3 s_2, \nonumber \\
& & 
2 s_1 + s_2  &\xrightarrow[]{ k_5 } s_1 + s_2,
& 3 s_2  &\xrightarrow[]{ k_{11} } 2 s_2, \nonumber \\
& & s_1 + 2 s_2  &\xrightarrow[]{ k_6 } 2 s_1 + 2 s_2,  \label{eq:inputnet2}
\end{align} 
where species $s_1$ and $s_2$ react according to the eleven reactions with rate coefficients $k_1, k_2, \ldots, k_{11} \ge 0$. We denote the copy-numbers of species $s_1$, and $s_2$, at time $t$ by $X_1(t)$, and $X_2(t)$, respectively. It was established in~\cite{Me} that, for particular choices of the rate coefficients, the deterministic model of reaction network~(\ref{eq:inputnet2}), given as equation (SI17) in SI Text, exhibits exotic dynamics: it undergoes a homoclinic bifurcation, and displays a bistability involving a limit cycle and an equilibrium point. On the other hand, it is demonstrated in~\cite{Me2} that the stochastic model of~(\ref{eq:inputnet2}) is not necessarily sensitive to the deterministic bifurcation, and may effectively behave in a monostable manner. The latter point is demonstrated in Figure~\ref{fig:example2}(c), where we show in red numerically approximated $x_1$-solutions of (SI17), one initiated in the region of attraction of the equilibrium point, while the other of the limit cycle. For a comparison, we also show in blue a representative sample path generated by applying the Gillespie algorithm on~(\ref{eq:inputnet2}). It can be seen that the stochastic solution spends significantly more time near the deterministic equilibrium point. To gain a clearer picture, we display in Figures~\ref{fig:example2}(a), and~(b), the joint, and the $x_1$-marginal, stationary PMFs, respectively, underlying network~(\ref{eq:inputnet2}), which have been obtained numerically for the same parameter values as in Figure~\ref{fig:example2}(c). In~(b), one can notice that the PMF is bimodal, but the left peak, corresponding to the limit cycle, is significantly smaller than the right peak, which corresponds to the stable equilibrium point.

We now apply Algorithm~\ref{tab:thealgorithm} on network~(\ref{eq:inputnet2}) to achieve two goals. Firstly, we balance the sizes of the two peaks of the stationary PMF from Figure~\ref{fig:example2}(b), thereby forcing the stochastic system to spend comparable amounts of time at the two deterministic attractors. Secondly, we reverse the situation shown in Figure~\ref{fig:example2}(b), by making the left PMF peak significantly larger than the right one, thereby forcing the stochastic system to spend most of the time near the limit cycle. We could achieve the goals by introducing species $\bar{s}_1, \bar{s}_2$ into~(\ref{eq:inputnet2}), and using suitable basis zero-drift networks. We take a simpler approach, by mapping~(\ref{eq:inputnet2}) to $\tilde{\mathcal{R}}^1(s_1,s_2,\bar{s}_2) \cup \mathcal{R}_1^2(\bar{s}_2) \cup (\mathcal{R}_{0,C_2 - 10}^3(s_2,\bar{s}_2) \cup \mathcal{R}_{30,0}^3(s_2,\bar{s}_2))$, which is given by equation (SI18) in SI Text. For our purposes, only one of $\bar{s}_1$, $\bar{s}_2$ is sufficient, since the stochastic dynamics of $s_1$ and $s_2$ are coupled. We have chosen $\bar{s}_2$ for convenience, since $x_2$-state-space may be truncated at a lower value, $C_2 = 180$, than $x_1$-state-space (see also Figure~\ref{fig:example2} (a)). The $x_2$-component of the deterministic limit cycle satisfies $x_2 \in (10,30)$. Correspondingly, we introduce two zero-drift networks: $\mathcal{R}_{0,C_2 - 10}^3(s_2,\bar{s}_2)$, and $\mathcal{R}_{30,0}^3(s_2,\bar{s}_2)$, which redistribute the PMF from $x_2 \in [0,10]$, and from $x_2 \in [30,C_2]$, respectively, to the limit cycle region, $x_2 \in (10,30)$. We fix the scaled rate coefficient $K_{0,C_2-10}^2$ to a large value (so that the PMF is nearly zero for $x_2 \in [0,10]$), and vary the coefficient $K_{30,0}^2$, which redistributes the PMF from the deterministic equilibrium point to the limit cycle. Network $\mathcal{R}_1^2(\bar{s}_2)$ is necessary for the preservation of the deterministic dynamics of~(\ref{eq:inputnet2}) under the application of Algorithm~\ref{tab:thealgorithm}. 

In Figures~\ref{fig:example2}(d), and~(e), we show the joint, and $x_1$-marginal, stationary PMFs for an intermediate value of $K_{30,0}^2$, when the PMF is partially redistributed from $x_2 \in [30,C_2]$ to $x_2 \in (10,30)$, so that the two peaks in~(e) are of comparable sizes. In Figure~\ref{fig:example2}(f), we show a representative sample path, obtained by applying the Gillespie algorithm on network (SI18) from SI Text, together with the deterministic trajectories obtained by solving (SI17). One can notice that the stochastic system now spends significantly more time near the limit cycle, when compared to~(c). In Figures~\ref{fig:example2}(f)--(g), we show analogous plots, but for a sufficiently large value of $K_{30,0}^2$, when the PMF is almost completely redistributed from $x_2 \in [30,C_2]$ to $x_2 \in (10,30)$. Now, in contrast to Figures~\ref{fig:example2}(a)--(c), the PMF becomes essentially unimodal, and concentrated around the limit cycle. Let us note that the red trajectories from Figures~\ref{fig:example2}(f) and~(i) were generated by numerically solving the deterministic model of network~(\ref{eq:inputnet2}), given by (SI17). For our purposes, it is not necessary to solve the corresponding (stiff) deterministic model of network (SI18). The reason is that Algorithm~\ref{tab:thealgorithm} does not influence the deterministic equilibrium points of a given reaction network, regardless of the choice of the kinetic algorithm parameters. For example, while the deterministic limit cycle is not necessarily preserved for the algorithm parameters chosen in Figure~\ref{fig:example2}(i), the enclosed deterministic unstable focus is necessarily preserved. Thus, the blue sample path corresponds to noise-induced oscillations either near a deterministic limit cycle, or near a deterministic unstable focus.

\section{Summary}
In this paper, we have presented the noise-control algorithm, which is given as Algorithm~\ref{tab:thealgorithm}. The algorithm maps an input chemical reaction network to output networks, all under mass-action kinetics, by introducing appropriate additional species and reactions, such that the output networks satisfy the following two properties. 
Firstly, the output networks have the same deterministic model as the input network, in appropriate limits of some of the parameters (rate coefficients) introduced by the algorithm. Secondly, controllable state-dependent noise is introduced into the stochastic model of the output networks. Thus, Algorithm~\ref{tab:thealgorithm} may be used to control the intrinsic noise of a given reaction network under mass-action kinetics, while preserving the deterministic dynamics. Let us note that the asymptotic conditions for the algorithm parameters are necessary for preservation of the time-dependent deterministic solutions. However, the time-independent deterministic solutions (the deterministic equilibrium points), which capture important features of the deterministic dynamics, are preserved under the algorithm even if the asymptotic conditions are not satisfied.

The algorithm has been applied to a test problem, taking the form of the one-species production-decay system given by~(\ref{eq:inputnet}). Using analytical and numerical methods, we have shown that the additional intrinsic noise, introduced by the algorithm, may be used to favorably modify the stationary probability mass function at arbitrary points in the state-space, as demonstrated in Figure~\ref{fig:example1}. For example, in Figure~\ref{fig:example1}(b), the noise is added to the whole interior of the state-space, while in~(e) only at a single point, in both cases resulting in noise-induce bimodality. On the other hand, in Figure~\ref{fig:example1}(h), by adding the noise to specific points in the state-space, the network is redesigned to display noise-induced trimodality. As shown in Figures~\ref{fig:example1}(c), (f), (i), the blue stochastic trajectories display multistability, while the red deterministic ones remain monostable. 

The algorithm has also been applied to a more challenging problem, taking the form of the two-species system given by~(\ref{eq:inputnet2}), which, for the parameters taken in this paper, at the deterministic level displays a bistability involving an equilibrium point and a limit cycle~\cite{Me,Me2}. At the stochastic level, the system is significantly more likely to be found near the equilibrium point, as demonstrated in Figures~\ref{fig:example2}(a)--(c). We have used the algorithm to redesign network~(\ref{eq:inputnet2}), so that the stochastic system spends comparable amounts of time near the two attractors, as demonstrated in Figures~\ref{fig:example2}(d)--(f). The network was also redesigned to display noise-induced oscillations, which is shown in Figures~\ref{fig:example2}(g)--(i).

The controllable state-dependent noise is generated by Algorithm~\ref{tab:thealgorithm} using the zero-drift networks~(\ref{eq:zerodriftinterior}) and~(\ref{eq:zerodriftleftboundary}). Any nonnegative function, defined on a bounded discrete domain, may be represented by a linear combination of propensity functions induced by an appropriate union of the zero-drift networks. Thus, choosing suitable zero-drift networks, the algorithm may control the intrinsic noise at arbitrary points in the state-space of the stochastic dynamics of reaction networks. The cost of such a precision in nose-control is a larger number of reactants in the underlying zero-drift networks. However, while the high-molecular reactions introduced by the algorithm are more expensive to synthetize, they do not limit applicability of Algorithm~\ref{tab:thealgorithm} to synthetic biology. \changecolor{The reason for this is that such reactions may always be broken down into sets of up-to bi-molecular reactions, with asymptotically equivalent deterministic and stochastic dynamics~\cite{UNI,Me3}. In particular, a zero-drift network, involving reactions of order $(n + \bar{n})$, may be broken down into $2 (n + \bar{n}) - 2$ reactions of up-to second-order, which may be readily mapped to DNA-based physical networks}.

\changecolor{Algorithm~\ref{tab:thealgorithm} may constitute a qualitatively novel finding which will facilitate the progress of DNA computing~\cite{DNAComputing1}. In particular, a hybrid approach for constructing DNA-based reaction networks may be used: the deterministic model may be used to guide the construction of reaction networks, and then Algorithm~\ref{tab:thealgorithm} may be applied to favorably reprogram the intrinsic noise in the stochastic model, while preserving the mean-field behavior. The algorithm may be of critical importance when the synthetic networks involve species at low copy-numbers, since then the stochastic effects may play a significant role~\cite{Noise1,Noise2,SynthBio3,SynthBio4,Me2,Radek1,Radek2,Radek3}, uncontrollably contaminating the performance of the synthetic networks. In such circumstances, Algorithm~\ref{tab:thealgorithm} may be used for controlling the stochastic effects, enriching the DNA-based synthetic systems with novel, noise-induced functionalities.}

\section{Methods}
Let us consider the mass-action reaction network $\mathcal{R}$ given by
\begin{align}
\mathcal{R}(s_1, \ldots, s_N): \; \; \sum_{i = 1}^{N} c_{i j} s_i &\xrightarrow[]{ k_j}  \sum_{i = 1}^N c_{i j}' s_i, \, \, \, \,  j \in \{1, \ldots, M\}, \label{eq:reactionsmethods}
\end{align}
where $s_1, \ldots, s_N$ are the reacting species, $k_j$ the reaction rate coefficients, and $c_{i j}, c_{i j}'$ the stoichiometric coefficients. Let us denote by $\mathbf{c}_j, \mathbf{c}_j' \in \mathbb{N}_{0}^N$ the vectors of the stoichiometric coefficients of reaction $j$, and $\Delta \mathbf{x}_j = \mathbf{c}_j' - \mathbf{c}_j$. 

The \emph{deterministic model} of reaction network~(\ref{eq:reactionsmethods}) is given by the following system of ordinary-differential equations (ODEs), also known as the reaction rate equations~\cite{Janos1,David}:
\begin{align}
\frac{\mathrm{d} \mathbf{x}}{\mathrm{d} t}  & =  \sum_{j = 1}^M k_j  \mathbf{x}^{\mathbf{c}_j} \Delta \mathbf{x}_j, \, \, \, \, \, \, \, \, i \in \{1, \ldots, N\}. \label{eq:deterministicmodel}
\end{align}
Here, $\mathbf{x} = \mathbf{x}(t) \in \mathbb{R}_{\ge}^N$ is the vector of species concentrations, i.e. $x_i(t)$ is the concentration of species $s_i$ at time $t$, and $\mathbf{x}^{\mathbf{c}_j} \equiv \prod_{l = 1}^N x_l^{c_{l j}}$, with the convention that $0^0 \equiv 1$.

The \emph{stochastic model} of reaction network~(\ref{eq:reactionsmethods}) is given by the following system of difference-differential equations, also known as the chemical master equation (CME)~\cite{Janos1,David,VanKampen}:
\begin{align}
\frac{\partial}{\partial t} p(\mathbf{x},t)   =  \mathcal{L} p(\mathbf{x},t) & = \sum_{j} (E_{\mathbf{x}}^{-\Delta \mathbf{\mathbf{x}}_j} - 1) \big(\alpha_j(\mathbf{x}) p(\mathbf{x},t) \big). \label{eq:stochasticmodel}
\end{align}
Here, $p(\mathbf{x},t)$ is the probability mass function (PMF), i.e. the probability that the vector of copy-numbers $\mathbf{X} = \mathbf{X}(t) \in \mathbb{N}_0^N$ of species $s_1, \ldots, s_N$ at time $t$ is given by $\mathbf{x}$. Linear operator $\mathcal{L}$ is called the forward operator, and step operator $E_{\mathbf{x}}^{-\Delta \mathbf{\mathbf{x}}_j}$ is such that $E_{\mathbf{x}}^{-\Delta \mathbf{\mathbf{x}}_j} p(\mathbf{x},t) = p(\mathbf{x} - \Delta \mathbf{x}_j,t)$. Function $\alpha_j(\mathbf{x})$ is the propensity function~\cite{Janos1,David} of the $j$-th reaction from~(\ref{eq:reactionsmethods}), and is given by
\begin{align}
\alpha_j(\mathbf{x}) \, = \,  k_j \mathbf{x}^{\underline{\mathbf{c}_j}} & = k_j \prod_{l = 1}^N x_l^{\underline{c_{l j}}}, \label{eq:propensitymethods}
\end{align}
where $x_l^{\underline{c_{l j}}}$ denotes a falling factorial of $x_l$, i.e. $x_l^{\underline{c_{l j}}} \equiv x_l (x_l - 1) \ldots (x_l - c_{l j} + 1)$.

\section*{Acknowledgments} 
The authors would like to thank the Isaac 
Newton Institute for Mathematical Sciences, Cambridge, for support 
and hospitality during the programme ``Stochastic Dynamical Systems 
in Biology: Numerical Methods and Applications'', where work on this 
paper was undertaken. The authors would also like to thank John J. Tyson
(Department of Biology, Virginia Polytechnic Institute and State University, USA)
for a discussion on a possible realization of network~(\ref{eq:zerodrift11}) via a
bifunctional histidine kinase/phosphatase from~\cite{Tyson}.
This work was supported by EPSRC grant no 
EP/K032208/1. This work was partially supported by a grant from 
the Simons Foundation. Konstantinos C. Zygalakis was supported by the Alan Turing
Institute under the EPSRC grant EP/N510129/1. David F. Anderson would like to acknowledge the
NSF grant NSF-DMS-1318832, and Army Research Office grant W911NF-14-1-0401. Radek Erban would also 
like to thank the Royal Society for a University Research Fellowship.


\section*{Supplementary Information (SI) Text}

\subsection*{The Deterministic Dynamics of Network $\hat{\mathcal{R}}^1 \cup \mathcal{R}_1^2$ in the Limit $\mu \to 0$}
Let us analyse system (4) in the asymptotic limit $\mu \to 0$. It follows from the Tikhonov theorem~[42] that the ODE for $y$, given by second equation in (4), reduces to the algebraic equation $y = (c - x)^{-1}$ as $\mu \to 0$. Substituting the algebraic equation into (4) results in 
\begin{align}
\frac{\mathrm{d} x}{\mathrm{d} t}  & =  k_1 - k_2 x, \nonumber \\
x(0) & = x_0, \, \, \, \text{as } \mu \to 0. \tag{SI1}\label{eq:outputODEsTikhonov} 
\end{align}
Initial value problems~(2) and~(\ref{eq:outputODEsTikhonov}) have the same form, and let us denote their solutions by $\hat{x}(t; \, \hat{x}_0)$ and $x(t; \, x_0)$, respectively. Then, choosing $c \ge \textrm{max}_{t \ge 0} \hat{x}(t; \, \hat{x}_0) < \infty$, and $x_0 = \hat{x}_0$, ensures that concentration of auxiliary species $\bar{s}$ is nonnegative, $\bar{x}(t) = c - x(t) \ge 0$, and that the solutions of (2) and (4) are asymptotically equivalent in the limit $\mu \to 0$. 

\subsection*{The Stochastic Dynamics of Network $\hat{\mathcal{R}}^1 \cup \mathcal{R}_1^2 \cup \mathcal{R}_{1,1}^3$ in the Limit $\mu \to 0$}
The chemical master equation (CME) [27] induced by network $\hat{\mathcal{R}}^1 \cup \mathcal{R}_1^2 \cup \mathcal{R}_{1,1}^3$ is given by
\begin{align}
\frac{\partial}{\partial t} p(x,y,t) & = \left(\mathcal{L}^1 + \frac{1}{\mu} \mathcal{L}_1^2 + K_{1,1} \mathcal{L}_{1,1}^3 \right) p(x,y,t), \tag{SI2}\label{eq:outputCME}
\end{align}
where $x(t),y(t) \in \mathbb{N}_0$ are copy-numbers of species $s,I^1$ from (3), respectively, with 
\begin{align}
\mathcal{L}^1 & = k_1 (E_{x}^{-1} - 1) \left( (C-x) y \right) + k_2 (E_{x}^{+1} - 1) x, \nonumber \\
\mathcal{L}_1^2 & = (E_{y}^{-1} - 1)  + (C-x) (E_{y}^{+1} - 1) y, \nonumber \\
\mathcal{L}_{1,1}^3 & =  (E_{x}^{-1} + E_{x}^{+1} - 2) \beta_{1,1}(x), \tag{SI3}\label{eq:outputoperators}
\end{align}
and $K_{1,1}, \beta_{1,1}(x)$ given in (8). Operators $\mathcal{L}^1, \mathcal{L}_1^2, \mathcal{L}_{1,1}^3$ are induced by subnetworks $\hat{\mathcal{R}}^1, \mathcal{R}_1^2, \mathcal{R}_{1,1}^3$, respectively. 

Let us analyse system~(\ref{eq:outputCME}) in the limit $\mu \to 0$, and consider the following power-series expansion:
\begin{align}
p(x,y,t) & = p_0(x,y,t)  + \mu p_1(x,y,t)  + \ldots \nonumber \\
            & + \mu^i p_i(x,y,t) + \ldots, \tag{SI4}\label{eq:powerseries1}
\end{align}
with $i \ge 2$.
Substituting~(\ref{eq:powerseries1}) into~(\ref{eq:outputCME}), and equating terms of equal powers in $\mu$, the following system of equations is obtained:
\begin{align}
\mathcal{O} \left(\frac{1}{\mu} \right): \; - \mathcal{L}_1^2 p_0(x,y,t)  & = 0, \nonumber \\
\mathcal{O}(1): \; - \mathcal{L}_1^{2} p_1(x,y,t)  & = (\mathcal{L}^1 + K_{1,1} \mathcal{L}_{1,1}^{3} \nonumber \\
& - \frac{\partial}{\partial t}) p_0(x,y,t) . \tag{SI5}\label{eq:averaging1}
\end{align}
\emph{Order $1/\mu$ equation}. A suitable form of the zero-order approximation of the PMF follows from the Bayes theorem: $p_0(x,y,t) =  p_0(y|x) p_0(x,t)$, where $p_0(y|x)$ is the stationary PMF of $y$ conditional on $x$, while $p_0(x,t)$ is the marginal PMF of $x$. Substituting $p_0(x,y,t) =  p_0(y|x) p_0(x,t)$ into the first equation in~(\ref{eq:averaging1}), with $t, x$ fixed, leads to $- \mathcal{L}_1^{2} p_0(y|x) = 0$. It follows that $p_0(y|x)$ is a Poisson distribution with parameter $(C-x)^{-1}$, so that the zero-order PMF is given by
\begin{align}
p_0(x,y,t) & = \left(\frac{1}{y!} \left(\frac{1}{(C-x)} \right)^{y}\exp \left(-\frac{1}{(C-x)} \right) \right) p_0(x,t). \tag{SI6}\label{eq:zeroorder}
\end{align}

\emph{Order $1$ equation}. Substituting~(\ref{eq:zeroorder}) into the second equation in~(\ref{eq:averaging1}), 
summing over all the possible states $y \in \mathbb{N}_0$, using~(\ref{eq:outputoperators}), and equalities $\sum_{y} y p_0(y|x) = (C-x)^{-1}$ and $\sum_{y} p_0(y|x)  = 1$, one obtains the \emph{effective CME}, given by
\begin{align}
 \frac{\partial}{\partial t} p_0(x,t) & = \left(\mathcal{L} +  K_{1,1} \mathcal{L}_{1,1}^{3} \right) p_0(x,t), \tag{SI7}\label{eq:effectiveCME}
\end{align}
where $\mathcal{L}$ is the forward operator corresponding to network (1), and has the following form
\begin{align}
 \mathcal{L} & = k_1 (E_{x}^{-1} - 1) + k_2 (E_{x}^{+1} - 1) x. \tag{SI8}\label{eq:inputoperator}
\end{align}

\subsubsection*{Limit $K_{1,1} \to 0$}
Setting the left-hand side (LHS) to zero, and taking $K_{1,1} = 0$ in~(\ref{eq:effectiveCME}), and assuming $C$ is fixed to a sufficiently large value, it follows that the stationary PMF is a Poisson distribution with parameter $k_1/k_2$ [27]:
\begin{align}
p_0(x) & = \begin{cases}
\frac{1}{x!} \left( \frac{k_1}{k_2}\right)^{x} \exp \left(- \frac{k_1}{k_2}\right), & \text{if } x \in [0,C], \\
0, & \text{otherwise} \tag{SI9}\label{eq:inputstationaryPMFS}.
\end{cases} 
\end{align}

\subsubsection*{Limit $K_{1,1} \to \infty$}
Let us substitute the power-series expansion
\begin{align}
p_0(x) & = f_0(x)  + \frac{1}{K_{1,1}} f_1(x)  + \ldots \nonumber \\
            & + \left(\frac{1}{K_{1,1}} \right)^i f_i(x) + \ldots, \tag{SI10}\label{eq:powerseries2}
\end{align}
with $i \ge 2$, into~(\ref{eq:effectiveCME}) with the LHS set to zero, and consider the limit $K_{1,1} \to \infty$. Then, equating terms of equal powers in $1/K_{1,1}$, one obtains:
\begin{align}
\mathcal{O} \left(1 \right): \; - \mathcal{L}^3 f_0(x)  & = 0, \nonumber \\
\mathcal{O} \left(\frac{1}{K_{1,1}} \right): \; - \mathcal{L}_{1,1}^3 f_1(x)   & = \mathcal{L} f_0(x) . \tag{SI11}\label{eq:averaging3}
\end{align}
\emph{Order $1$ equation}. The solution to the first equation in~(\ref{eq:averaging3}) is given by
\begin{align}
f_0(x) & = \begin{cases}
1 - \frac{a}{C}, & \text{if } x = 0, \\
\frac{a}{C}, & \text{if } x = C, \\
0, & \text{otherwise},  \tag{SI12}\label{eq:f0}
\end{cases} 
\end{align}
where $a \in \mathbb{R}_{\ge}$ is an arbitrary constant.

\emph{Order $1/K_{1,1}$ equation}. Multiplying the second equation in~(\ref{eq:averaging3}) by $x$, and summing over $x \in \mathbb{N}_0$, with the convention that $f_0(x) = 0$ and $\beta_{1,1}(x) = 0$ for $x \notin [0,C]$, one obtains the solvability condition $0 = \sum_{x = 0}^{\infty} x \mathcal{L} f_0(x)$, which implies $a = k_1/k_2$. Substituting $a$ into~(\ref{eq:f0}) leads to the zero-order approximation of the stationary PMF:
\begin{align}
f_0(x) & = \begin{cases}
1 - \frac{1}{C} \frac{k_1}{k_2}, & \text{if } x = 0, \\
\frac{1}{C} \frac{k_1}{k_2}, & \text{if } x = C, \\
0, & \text{otherwise}. \tag{SI13}\label{eq:f02S}
\end{cases} 
\end{align}

\subsection*{Zero-Drift Networks $\mathcal{R}_{n,\bar{n}}^3$}
The propensity function of reactions underlying $\mathcal{R}_{n,\bar{n}}^3(s,\bar{s})$,  $n,\bar{n} \in \mathbb{N}_0$, and $(n + \bar{n}) \le C$, is given by $K_{n,\bar{n}} \beta_{n,\bar{n}} : [0,C] \to \mathbb{R}_{\ge}$, with
\begin{align}
K_{n,\bar{n}} = M_{n,\bar{n}} k_{n,\bar{n}}, \tag{SI14}\label{eq:zerodriftcoefficient}
\end{align}
and
\begin{align}
\beta_{n,\bar{n}}(x) & = (M_{n,\bar{n}})^{-1} \prod_{i = 0}^{n - 1}{\left( x - i \right)} \prod_{i = 0}^{\bar{n} - 1}{\left((C - i) - x \right)}, \tag{SI15}\label{eq:zerodriftpropensity}
\end{align}
where the scaling factor $M_{n,\bar{n}}$ is introduced to approximately normalize $\beta_{n,\bar{n}}(x)$, and is given by
\begin{align}
M_{n,\bar{n}} & = \prod_{i = 0}^{n - 1}{\left( \frac{n}{n + \bar{n}} C - i \right)} \prod_{l = 0}^{\bar{n} - 1}{\left(\frac{\bar{n}}{n + \bar{n}} C - i \right)}. \tag{SI16}\label{eq:scaling}
\end{align}
Here, we take the convention $\prod_{i = 0}^N f(i) = 1$ if $N < 0$, where $f(i)$ is an arbitrary function of $i$. Function $\beta_{n,\bar{n}}(x)$ is nonzero on the interval $[n,C-\bar{n}]$, with the single maximum approximately at $C n/(n + \bar{n})$.

\emph{Interior zero-drift networks}. Zero-drift network $\mathcal{R}_{n,\bar{n}}^3(s,\bar{s})$, with $n, \bar{n} \ne 0$, satisfies~(19), and the propensity function of its reactions, which is proportional to~(\ref{eq:zerodriftpropensity}), is nonzero only in the interior of the state-space. Since the propensity function of $\mathcal{R}_{n,\bar{n}}^3(s,\bar{s})$, with $n, \bar{n} \ne 0$, attains its maximum in the interior of the domain, we call the network an interior zero-drift network.

\emph{Boundary zero-drift networks}. Network $\mathcal{R}_{0,\bar{n}}^3(s,\bar{s})$, satisfying~(20), is a zero-drift network in the limit $\mu_{0,\bar{n}} \to 0$. Furthermore, in the same limit, the first two reactions from~(20) have the same propensity function, which is proportional to~(\ref{eq:zerodriftpropensity}) with $n = 0$, and which is nonzero at the left boundary point, $x = 0$. Similarly, network $\mathcal{R}_{n,0}^3 = \mathcal{R}_{0,n}^3(\bar{s}, s; \, \bar{B}, k_{n,0}, \mu_{n,0})$ is a zero-drift network as $\mu_{n,0} \to 0$, and its first two reactions have the same propensity function, which is nonzero at the right boundary point, $x = C$. Since networks with $n = 0$ (respectively, $\bar{n} = 0$) generate propensity functions with the maximum values at the left (respectively, right) boundary point, we call such networks left (respectively, right) boundary zero-drift networks.

\emph{Basis zero-drift networks}. Stoichiometric coefficients $n, \bar{n}$ control the support of the intrinsic noise, which network $\mathcal{R}_{n,\bar{n}}^3$ introduces into the stochastic dynamics, via the control of support of the compact function~(\ref{eq:zerodriftpropensity}). The larger the sum $(n + \bar{n})$ is, with $(n + \bar{n}) \le C$, the smaller the support of~(\ref{eq:zerodriftpropensity}), and, hence, one obtains a more precise noise-control. In the special case when $n + \bar{n} = C$, the propensity function~(\ref{eq:zerodriftpropensity}) is nonzero only at a single point in the state-space, $x = n$. We call networks $\mathcal{R}_{n,\bar{n}}^3(s,\bar{s})$, with $n + \bar{n} = C$, basis zero-drift networks, and the corresponding propensity functions \emph{basis propensity functions}. Any nonnegative function, defined on a bounded discrete domain, may be represented by a suitable linear combination of the basis propensity functions. 

\subsection*{The Deterministic Model for Network $\tilde{\mathcal{R}}$}
The deterministic model of network~(11) is given by
\begin{align}
\frac{\mathrm{d} x_1}{\mathrm{d} t}  
& = 
k_1 + k_2 x_1 + k_3 x_1^2 - k_4 x_1 x_2 - k_5 x_1^2 x_2 + k_6 x_1 x_2^2,  
\nonumber \\
\frac{\mathrm{d} x_2}{\mathrm{d} t} 
& = 
k_7  - k_8 x_2 + k_9 x_1 x_2 + k_{10} x_2^2 - k_{11} x_2^3,
\tag{SI17}\label{eq:inputODEs2}
\end{align}
where $x_1 = x_1(t), x_2 = x_2(t)$ are the concentrations of species $s_1, s_2$, respectively, at time $t$.

\subsection*{Applying Algorithm 1 on Network $\tilde{\mathcal{R}}$}
Network $\tilde{\mathcal{R}}^1(s_1,s_2,\bar{s}_2) \cup \mathcal{R}_1^2(\bar{s}_2) \cup (\mathcal{R}_{0,C_2 - 10}^3(s_2,\bar{s}_2) \cup \mathcal{R}_{30,0}^3(s_2,\bar{s}_2))$ is given by
\begin{align}
\tilde{\mathcal{R}}^1(s_1,s_2,\bar{s}_2): 
& & \varnothing  & \xrightarrow[]{ k_1 } s_1, \nonumber \\
& & s_1  & \xrightarrow[]{ k_2 } 2 s_1, \nonumber \\
& & 2 s_1  &\xrightarrow[]{ k_3 } 3 s_1, \nonumber \\
& & s_1 + s_2  &\xrightarrow[]{ k_4 } s_2, \nonumber \\
& & 2 s_1 + s_2  &\xrightarrow[]{ k_5 } s_1 + s_2, \nonumber \\
& & s_1 + 2 s_2  &\xrightarrow[]{ k_6 } 2 s_1 + 2 s_2, \nonumber \\
& & \bar{s}_2 + I_2^1 &\xrightarrow[]{ k_7 } s_2 + I_2^1, \nonumber \\
& & s_2  &\xrightarrow[]{ k_8 } \bar{s}_2, \nonumber \\
& & s_1 + s_2 + \bar{s}_2 + I_2^1 &\xrightarrow[]{ k_9 } s_1 + 2 s_2 + I_2^1, \nonumber \\
& & 2 s_2 + \bar{s}_2 + I_2^1  &\xrightarrow[]{ k_{10} } 3 s_2 + I_2^1, \nonumber \\ 
& & 3 s_2  &\xrightarrow[]{ k_{11} } 2 s_2 + \bar{s}_2, \nonumber \\
\mathcal{R}_{1}^2(\bar{s}_2): 
& & \varnothing &\xrightarrow[]{ 1/\mu}  I_2^1, \nonumber \\
& &  \bar{s}_2 + I_2^1 &\xrightarrow[]{1/\mu}  \bar{s}_2, \nonumber \\
\mathcal{R}_{0,C_2-10}^3(s_2,\bar{s}_2): 
& & (C_2-10) \bar{s}_2 & \xrightarrow[]{k_{0,C_2-10}^2} s_2 + (C_2 - 11) \bar{s}_2, \nonumber \\
& & C_2 s_2 + B_2 & \xrightarrow[]{k_{0,C_2-10}^2} (C_2 - 1) s_2 + \bar{s}_2 + B_2, \nonumber \\
& & (C_2-10) \bar{s}_2 & \xrightarrow[]{1/\mu_{0,C_2-10}} (C_2-10) \bar{s}_2 + B_2, \nonumber \\
& & C_2 s_2 + B_2 & \xrightarrow[]{1/\mu_{0,C_2-10}} C_2 s_2, \nonumber \\
\mathcal{R}_{30,0}^3(s_2,\bar{s}_2): 
& & 30 s_2 & \xrightarrow[]{k_{30,0}^2} 29 s_2 + \bar{s}_2, \nonumber \\
& & C_2 \bar{s}_2 + \bar{B}_2 & \xrightarrow[]{k_{30,0}^2} s_2 + (C_2 - 1) \bar{s}_2 + \bar{B}_2, \nonumber \\
& & 30 s_2 & \xrightarrow[]{1/\mu_{30,0}} 30 s_2 + \bar{B}_2, \nonumber \\
& & C_2 \bar{s}_2 + \bar{B}_2 & \xrightarrow[]{1/\mu_{30,0}} C_2 \bar{s}_2. \tag{SI18}\label{eq:outputnet2}
\end{align}

\begin{table*}[t]
 \renewcommand\tablename{Algorithm}
\hrule
\vskip 2.5 mm
\textbf{Input}: Let the input reaction network be given by
\begin{align}
\hat{\mathcal{R}}(s_1, \ldots, s_N): \; \; \sum_{i = 1}^{N} c_{i j} s_i &\xrightarrow[]{ k_j}  \sum_{i = 1}^N c_{i j}' s_i, \, \, \, \,  j \in \{1, \ldots, M\}, \label{eq:inputnetgeneral}
\end{align}
where $s_1, \ldots, s_N$, are the species, $k_j$ the reaction rate coefficients, and $c_{i j}, c_{i j}'$ the stoichiometric coefficients. 
\begin{enumerate}
\item [\textbf{(1)}] \textbf{Step}: Reaction network $\hat{\mathcal{R}}$, given by~(\ref{eq:inputnetgeneral}), is mapped to a \emph{pairwise conservative network} $\hat{\mathcal{R}}^1$ given by
\begin{align}
\hat{\mathcal{R}}^1(s_1, \ldots, s_N, \bar{s}_1, \ldots, \bar{s}_N): \; \;
\sum_{i = 1}^N \Big( c_{i j} s_i + (\Delta x_{i j} \bar{s}_i + I_{i}^{\Delta x_{i j}}) \times 1_{\mathbb{N}}(\Delta x_{i j}) \Big) \xrightarrow[]{k_j} \nonumber \\
\sum_{i = 1}^N \Big( c_{i j}' s_i - (\Delta x_{i j} \bar{s}_i) \times 1_{\mathbb{N}}(-\Delta x_{i j}) + I_{i}^{\Delta x_{i j}} \times 1_{\mathbb{N}}(\Delta x_{i j}) \Big), \, \, \, \, j \in \{1, \ldots, M\}. \label{eq:net1}
\end{align}
Here, $\bar{s}_i, I_{i}^{\Delta x_{i j}}$ are additional species, $\Delta x_{i j} = (c_{i j}' - c_{i j})$, and $1_{\mathbb{N}}(\cdot)$ is the indicator function of the natural numbers.
\item [\textbf{(2)}] \textbf{Step}: For each species $I_i^{\Delta x_{i j}}$, a \emph{drift-corrector network} is constructed, $\mathcal{R}_{\Delta x_{i j}}^2(\bar{s}_i) = \mathcal{R}_{\Delta x_{i j}}^2(\bar{s}_i; \, I_{i}^{\Delta x_{i j}}, \mu)$, given by
\begin{align}
 \mathcal{R}_{\Delta x_{i j}}^2(\bar{s}_i): \; \; \; \; \varnothing &\xrightarrow[]{ 1/\mu}  I_{i}^{\Delta x_{i j}}, \nonumber \\
  \; \; \; \; \Delta x_{i j} \bar{s}_i + I_{i}^{\Delta x_{i j}} &\xrightarrow[]{1/\mu}  \Delta x_{i j} \bar{s}_i. \label{eq:net3}
\end{align}
where $0 \le \mu \ll 1$.
\item [\textbf{(3)}] \textbf{Step}: For each species $\bar{s}_i$, a union of \emph{zero-drift networks} may be constructed. Let $n,\bar{n} \in \mathbb{N}_0$, and $(n + \bar{n}) \le C_i$. Network $\mathcal{R}_{n,\bar{n}}^3(s_i,\bar{s}_i) = \mathcal{R}_{n,\bar{n}}^3(s_i,\bar{s}_i; \, k_{n,\bar{n}}^i)$, with $n, \bar{n} \ne 0$, is given by
\begin{align}
\mathcal{R}_{n,\bar{n}}^3(s_i,\bar{s}_i): \; \; n s_i + \bar{n} \bar{s}_i & \xrightarrow[]{ k_{n,\bar{n}}^i}  (n + 1) s_i + (\bar{n} - 1) \bar{s}_i, \nonumber \\
\; \; n s_i + \bar{n} \bar{s}_i & \xrightarrow[]{ k_{n,\bar{n}}^i } (n - 1) s_i + (\bar{n} + 1) \bar{s}_i \label{eq:zerodriftinterior}.
\end{align}
Network $\mathcal{R}_{0,\bar{n}}^3(s_i,\bar{s}_i) = \mathcal{R}_{0,\bar{n}}^3(s_i,\bar{s}_i; \ B_i, k_{0,\bar{n}}^i, \mu_{0,\bar{n}})$, with $\bar{n} \ne 0$, is given by
\begin{align}
\mathcal{R}_{0,\bar{n}}^3(s_i,\bar{s}_i): \; \; \bar{n} \bar{s}_i & \xrightarrow[]{k_{0,\bar{n}}^i} s_i + (\bar{n} - 1) \bar{s}_i, \nonumber \\
\; \; C_i s_i + B_i & \xrightarrow[]{k_{0,\bar{n}}^i} (C_i - 1) s_i + \bar{s}_i + B_i, \nonumber \\
\; \; \bar{n} \bar{s}_i & \xrightarrow[]{1/\mu_{0,\bar{n}}} \bar{n} \bar{s}_i + B_i, \nonumber \\
\; \; C_i s_i + B_i & \xrightarrow[]{1/\mu_{0,\bar{n}}} C_i s_i, \label{eq:zerodriftleftboundary}
\end{align} 
where $0 \le \mu_{0,\bar{n}} \ll 1$, and $B_i$ is an additional species. Network $\mathcal{R}_{n,0}^3 = \mathcal{R}_{0,n}^3(\bar{s}_i, s_i; \, \bar{B}_i, k_{n,0}^i, \mu_{n,0})$.
\end{enumerate}
\textbf{Output}: An output reaction network $\mathcal{R}$ is given by
\begin{align}
\mathcal{R} & = \hat{\mathcal{R}}^1 \cup \mathcal{R}^2 \cup \mathcal{R}^3, \label{eq:outputnetgeneral}
\end{align}
where $\mathcal{R}^2 = \cup_{i} \cup_{\Delta x_{i j}} \mathcal{R}_{\Delta x_{i j}}^2(\bar{s}_i)$, and $\mathcal{R}^3 = \cup_{i} \cup_{(n,\bar{n})} \mathcal{R}_{n,\bar{n}}^3(s_i,\bar{s}_i)$. 
\smallskip
\hrule
\smallskip
\caption{{\it \noindent The noise-control algorithm.}}
\label{tab:thealgorithm}
\end{table*}

\end{document}